\documentclass[%
 reprint,
superscriptaddress,
 amsmath,amssymb,
 aps,
]{revtex4-1}

\usepackage{graphicx}
\usepackage{dcolumn}
\usepackage{bm}
\usepackage{hyperref}

\usepackage{caption} 

\begin{document}

\preprint{APS/123-QED}

\title{Bound Orbits and Gravitational Wave Radiation Around the Hairy Black Hole
}%

\author{Liping Meng}
\author{Zhaoyi Xu}

 \author{Meirong Tang}
 \email{Electronic address: tangmr@gzu.edu.cn(Corresponding author)
}
\affiliation{%
College of Physics, Guizhou University,\\
 Guiyang 550025, China
}%





\date{\today}

\begin{abstract}
The hairy black hole model provides a new theoretical framework for exploring phenomena in strong gravitational fields. This paper systematically investigates the influence of the hair parameter $\beta$ on the timelike geodesics of the regular hairy black hole, including the radius of the event horizon, the properties of bound orbits, and the characteristics of gravitational wave radiation over a single period. The study reveals that $\beta$ has a significant impact on the event horizon but only a minor effect on the innermost stable circular orbit(ISCO), the marginally bound orbit(MBO), and periodic orbits. Moreover, the trajectories of the periodic orbits are nearly identical to those of the Schwarzschild black hole. In addition, the parameter $\beta$ was constrained by simulating the precession observational data of the S2 star orbiting the supermassive black hole Sgr A*. The results indicate that the correction effects of $\beta$ comply with existing observational constraints, without providing stricter limitations. Furthermore, by considering periodic orbits as transitional orbits in the extreme-mass-ratio inspiral (EMRI) system, it is found that the presence of $\beta$ introduces subtle effects on the amplitude, phase, and period of the gravitational wave signal for a single orbit. Although these effects appear minor within a single cycle, they may accumulate into significant effects over long-term evolution. In the future, space-based gravitational wave detectors are expected to further investigate the properties of the hair parameter, enhancing our understanding of the spacetime structure and dynamical behavior of black holes.
\begin{description}
\item[Keywords]
Periodic orbits, EMRI system, Hairy black holes, Gravitational wave
\end{description}
\end{abstract}

\maketitle

\section{Introduction}\label{1.0}
General relativity provides a theoretical foundation for exploring the universe and has successfully predicted the existence of black holes\cite{EventHorizonTelescope:2019dse,EventHorizonTelescope:2019ths,EventHorizonTelescope:2019pgp,EventHorizonTelescope:2019ggy}. The study of black hole physics originated from the pioneering work of J.R. Oppenheimer and H. Snyder\cite{Oppenheimer:1939ue}. Since then, black holes have become a focal point in the fields of general relativity and cosmology. The properties and behavior of black holes can be described through black hole solutions, which are a class of exact solutions derived from Einstein's field equations under specific conditions. The classical black hole solutions include the Schwarzschild solution\cite{Frolov1998,Wald:1984rg}, the Kerr solution\cite{2018grav.book.....M,Kerr:1963ud}, the Reissner-Nordström (RN) solution\cite{Reissner:1916cle,Graves:1960zz,1918KNAB...20.1238N}, and the Kerr-Newman solution\cite{Newman:1965my,Carter:1968rr}. In the above black hole solutions, black holes are completely described solely by their mass $ M $, angular momentum $ J $, and charge $ Q$, adhering to the no-hair theorem\cite{Bekenstein:1971hc,Ruffini:1971bza}. However, the appearance of the BBMB black hole challenged the no-hair theorem. First proposed by Bocharova, Bronnikov, Melnikov, and Bekenstein\cite{Bekenstein:1974sf}, it was the first exact black hole solution found to include a scalar field and represents a black hole only in four-dimensional space\cite{Xanthopoulos:1992fm}. Subsequently, within the framework of Einstein-Yang-Mills theory, Piotr Bizon was the first to obtain a “colored” black hole solution using numerical methods\cite{Bizon:1990sr}. These discoveries have encouraged researchers to explore more hairy black hole solutions (see references such as \cite{Contreras:2021yxe,Elvang:2004rt,Emparan:2001wn,Ovalle:2023ref}). The emergence of hairy black holes challenges the traditional information paradox\cite{Hawking:1976de}, providing important clues for understanding the formation and evolution of black holes. Therefore, further research on hairy black holes will help characterize the physical properties of black holes more precisely and deepen our understanding of black hole behavior. These studies will further clarify the role of black holes in cosmic structure and astrophysical evolution, providing a more reliable scientific foundation for the development of astrophysics and black hole theory.

Meanwhile, the extreme-mass-ratio inspiral (EMRI) system is also an important field for studying black hole properties\cite{Amaro-Seoane:2007osp}. This system consists of a stellar-mass compact object (such as a neutron star, white dwarf, or black hole) and a supermassive black hole\cite{Glampedakis:2005hs}, with a typical mass ratio of $10^{-4} $ to $ 10^{-7} $. With the successful detection of gravitational waves from binary black hole mergers\cite{LIGOScientific:2016aoc,LIGOScientific:2016vlm}, the era of gravitational wave astronomy has officially begun. This breakthrough not only advances research on black holes and their merger mechanisms but also makes the EMRI system particularly important in astronomy and gravitational wave studies, as the EMRI system is considered one of the key signal sources in gravitational wave observations\cite{Amaro-Seoane:2007osp,Babak:2017tow}. Additionally, the EMRI system holds research value in other areas, such as testing for the presence of dark matter around massive black holes\cite{Yue:2018vtk,Yue:2017iwc,Hannuksela:2019vip}, constraining cosmological parameters\cite{Schutz:1986gp}, and verifying general relativity\cite{Amaro-Seoane:2019umn}.

It is well known that periodic orbits exist around black holes\cite{Levin:2008mq,Grossman:2008yk,Levin:2008ci}. Moreover, in the EMRI system, periodic orbits can radiate gravitational waves\cite{Levin:2008mq,Glampedakis:2002ya}. By observing these gravitational wave waveforms, more information can be obtained about the orbital dynamical characteristics of stellar-mass celestial bodies and the formation of supermassive black holes and the basic mechanisms of the universe\cite{Babak:2017tow}. This prompts us to conduct a preliminary exploration of the periodic orbits of a timelike test particle around a regular hairy black hole and the gravitational waves radiated by the EMRI system. It is worth noting that bound orbits around black holes are divided into two types. One type is the precessing orbit, whose study began with one of the key cases in testing general relativity: the precession of Mercury’s perihelion\cite{Will:1993hxu}. Subsequently, research on the precessing orbits of stars around Sgr A* has also received extensive attention and in-depth discussion\cite{Iorio:2011zi,GRAVITY:2019tuf}. These investigations have found valuable applications in constraining theories of gravity\cite{Hees:2017aal,DellaMonica:2023dcw}.The other type of orbit is the periodic orbit, which plays an important role in the study of gravitational wave radiation\cite{Glampedakis:2002ya}. It is widely believed that periodic orbits can provide more information about the fundamental nature of black holes. Therefore, research on periodic orbits is of great significance for understanding the formation mechanisms of black holes and the properties of particle trajectories around them.

In light of this, Janna Levin and others proposed a classification method for periodic orbits\cite{Levin:2008mq}. Each orbit is represented by three integers $ z $, $ w $, and $ v $, which respectively denote the scaling, rotation, and vertex behavior of the orbit, thereby defining the quantity $ q = \frac{w + v}{z} $. When $ q $ is an irrational number, the orbit is a precessing orbit, where each subsequent orbit undergoes a precession relative to the previous one\cite{Grould:2017bsw,DeLaurentis:2018ahr}. When $ q $ is a rational number, the orbit is periodic, and a particle undergoing periodic orbital motion will, after a finite number of repetitions, return precisely to its initial state. Notably, the precessing orbit of a particle can be seen as a perturbation of a periodic orbit\cite{Levin:2008mq}. Therefore, studying periodic orbits can reveal properties related to precessing orbits and provide physical insights into black holes. Currently, this classification method has been extensively studied in various types of black holes, including Kerr black holes\cite{Levin:2008yp,Perez-Giz:2008ajn,Grossman:2011ps}, Schwarzschild black holes\cite{Lim:2024mkb}, Einstein-Lovelock black holes\cite{Lin:2021noq}, binary black hole systems\cite{Healy:2009zm}, charged black holes\cite{Misra:2010pu}, spherically symmetric naked singularities\cite{Babar:2017gsg}, and polymer black holes\cite{Tu:2023xab}.

Finally, studying hairy black holes is of great significance for understanding the properties and physical characteristics of black holes. The proposal of the hairy black hole model breaks the standard no-hair theorem, allowing black holes to have more characteristic parameters\cite{Gervalle:2024yxj,Astorino:2013sfa}. Therefore, hairy black holes can provide more physical information about black holes. In this case, the study of periodic orbits is particularly important because they can more sensitively capture the physical effects brought by these additional hairy parameters\cite{Lin:2023rmo,Collodel:2021gxu}. In particular, in EMRI systems, timelike test particles moving along the periodic orbits of hairy black holes will exhibit unique signal characteristics in gravitational wave radiation, which has important observational significance for the verification and expansion of the no-hair theorem. In addition, the EMRI system itself is one of the important targets of future gravitational wave detection missions (such as the Laser Interferometer Space Antenna\cite{Armano:2016bkm}, Taiji\cite{Hu:2017mde}, and TianQin\cite{TianQin:2015yph}). The gravitational wave signals generated by the system help finely detect the spacetime structure around black holes, thus distinguishing a regular hairy black hole from the traditional Schwarzschild or Kerr black hole. Therefore, studying the periodic orbits and gravitational wave characteristics in the context of hairy black holes can not only deepen our understanding of black holes but also provide a rich theoretical support and data interpretation framework for future gravitational wave observations.

In this paper, we mainly discuss the periodic orbits of a timelike test particle in the background of a regular hairy black hole and preliminarily study the characteristics of gravitational waves generated by periodic orbits in the EMRI system in this background. The structure of this paper is as follows: In Section \ref{2.0}, we give a brief introduction to the regular hairy black hole and calculate the effective potential of a timelike test particle around it. In Section \ref{3.0}, in the first subsection, we introduce the relevant properties of MBO and ISCO; in the second subsection, we introduce precession and periodic orbits. In Section \ref{4.0}, we conduct a preliminary exploration of the gravitational waves radiated by periodic orbits. Finally, in Section \ref{5.0}, a summary is given. In this paper, unless otherwise specified, the natural unit system with $c = G = 1$ is used throughout.

\section{The regular hairy black hole}\label{2.0}
Wheeler and others proposed the no-hair theorem for black holes. That is, within the framework of general relativity, all the properties of a black hole are determined only by the mass $M$, angular momentum $J$, and electric charge $Q$ of the black hole\cite{Bekenstein:1971hc,Ruffini:1971bza}. However, in one case, when there are various nonlinear matter fields outside the black hole, the no-hair theorem will no longer apply\cite{Nucamendi:1995ex,Zloshchastiev:2004ny}. In another case, under the Einstein-Yang-Mills theory, black holes can exhibit additional hairs\cite{Winstanley:1995iq,Mavromatos:1997zb}. Among them, Piotr Bizon was the first to analyze the static spherically symmetric Einstein-Yang-Mills equation with the $SU(2)$ gauge group and obtained a “colored” black hole solution through numerical methods\cite{Bizon:1990sr}. This discovery has promoted researchers' exploration of more hairy black hole solutions. For example, hairy black hole solutions in multi-dimensional spacetimes\cite{Emparan:2001wn,Elvang:2004rt,Bena:2005ay,Jejjala:2005yu} and some hairy black hole solutions constructed by scholars using the method of gravitational decoupling\cite{Contreras:2021yxe,Ovalle:2020kpd,Ovalle:2018umz,Avalos:2023ywb,Ovalle:2023ref}. 

In this paper, we briefly review the black hole solutions obtained by Jorge Ovalle and other scholars using the gravitational decoupling analysis method. In their calculations, they introduced the tensor vacuum $\theta_{\mu\nu}$, an additional energy-momentum tensor used to describe a new matter field. By applying geometric deformations to the traditional Schwarzschild solution, this approach effectively avoids the central singularity problem (The necessary formula derivations are provided in Appendix.). Under the conditions of satisfying the weak energy condition and the event horizon constraint, the authors ultimately derived a regular hairy black hole solution, with its metric given as follows\cite{Ovalle:2023ref}
\begin{equation}\label{1}
ds^2=-f(r)dt^2+{f(r)}^{-1}dr^2+r^2d\theta^2+r^2\sin^2{\theta}d\phi^2,
\end{equation}
\begin{equation}\label{2}
f(r)=1-\frac{2M}{r}+\frac{e^{-\alpha r/M}}{rM}(\alpha^2r^2+2M\alpha r+2M^2).
\end{equation}
Here, $M$ represents the ADM mass of the black hole, and the parameter $ \alpha $ is the hair parameter, which represents the source that causes the deformation of the Minkowski vacuum. As $\alpha \to 0 $, the metric degenerates into the flat Minkowski spacetime, indicating that there is no gravitational field and the spacetime returns to a flat structure without gravity. On the other hand, as $ \alpha \to \infty $, the metric converges to the static Schwarzschild solution, exhibiting gravitational characteristics similar to those of a classical black hole. In this process, the variation of $ \alpha $ causes the spacetime geometry to first transition from flat spacetime to a regular hairy black hole solution with a hair structure, and further transition to the classical Schwarzschild black hole solution. Therefore, the existence of the regular hairy black hole is closely related to the value of $ \alpha $, indicating that this parameter plays a decisive role in the evolution of the gravitational field and spacetime structure. Moreover, this solution has a regular spacetime structure and does not have a central singularity, thus overcoming the singularity problem of the traditional Schwarzschild black hole.

\begin{figure*}[]
\includegraphics[width=1\textwidth]{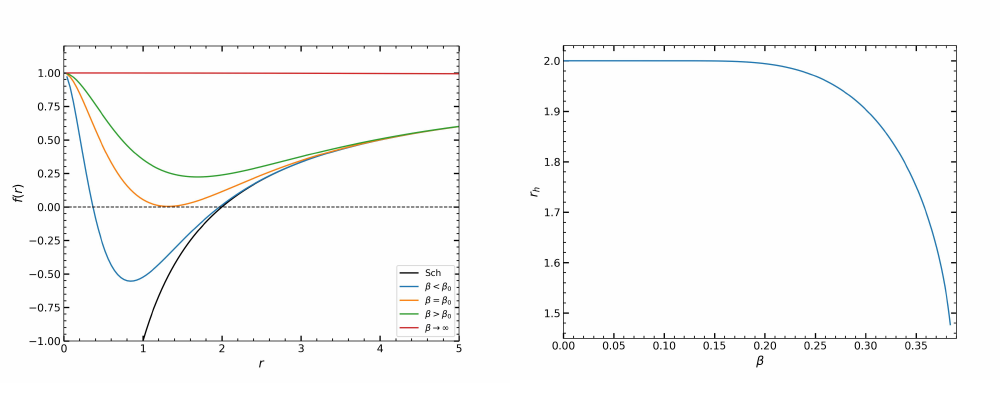}
\caption{The information on the event horizon of a regular hairy black hole is depicted. In the left figure, the parameter $\beta_0 = 0.3906$ represents an extremal black hole, where the black curve corresponds to a Schwarzschild black hole. The right figure shows the variation of the event horizon with changes in the hair parameter. Here, $M = 1$.}
\label{a}
\end{figure*}

To better describe the deviation of the regular hairy black hole from the Schwarzschild black hole, we use $\beta$ to replace the hair parameter $\alpha$, where $\beta = \frac{1}{\alpha}$. Thus, the metric \eqref{2} is rewritten as follows
\begin{equation}\label{3}
f(r)=1-\frac{2M}{r}+\frac{e^{-r/\beta M}}{rM}(\frac{r^2}{\beta^2}+\frac{2Mr}{\beta}+2M^2).
\end{equation}
In this case, when $\beta\rightarrow0$, the black hole solution reverts to the Schwarzschild solution $(\alpha \to \infty)$; when $\beta\rightarrow\infty$, the Minkowski spacetime is obtained $(\alpha \to 0)$. In Fig.\ref{a}, the event horizon information of a regular hairy black hole ($f(r)=0$) can be obtained. Obviously, in the left figure, when $\beta<0.3906$, there is a black hole with two event horizons; when $\beta_0 = 0.3906$, there is an extremal black hole with only one event horizon (orange solid line); when $\beta>0.3906$, there is a spacetime without a black hole. When $\beta\rightarrow\infty$, this corresponds to an asymptotically flat spacetime, which is represented by the red solid line in the figure. In the right figure, it can be clearly observed that as the hair parameter increases, the radius of the black hole's event horizon gradually decreases. This change becomes particularly significant as it approaches the extremal black hole. This indicates that the hair parameter has a significant impact on the geometric properties of the black hole's event horizon. Furthermore, the introduction of the hair parameter results in the event horizon radius of the hairy black hole being significantly smaller than that of the Schwarzschild black hole $(\beta = 0)$.

Next, to further explore the spacetime structure around the black hole and the dynamical behavior of particles, we will focus on the timelike geodesics around a regular hairy black hole. To simplify the analysis, the study will be limited to the motion of neutral particles. For the timelike geodesics in the EMRI system, considering the motion of a test particle with stellar mass around a regular hairy black hole, on the equatorial plane ($\theta=\frac{\pi}{2},\dot{\theta}=0$), the Lagrangian of the test particle is
\begin{equation}\label{4}
\mathcal{L}=\frac{1}{2}mg_{\mu\nu}{\dot{x}}^\mu{\dot{x}}^\nu=\frac{1}{2}m(-f(r){\dot{t}}^2+f(r)^{-1}{\dot{r}}^2+r^2{\dot{\phi}}^2).
\end{equation}
Among them, the “$\cdot$” here represents the derivative with respect to proper time $\tau$. $m$ represents the mass of the test particle, and in the subsequent discussion, it is set that $m = 1$. The generalized momentum of the test particle is defined by the following equation.
\begin{equation}\label{5}
P_\mu=\frac{\partial\mathcal{L}}{\partial{\dot{x}}^\mu}=g_{\mu\nu}{\dot{x}}^\nu.
\end{equation}
According to the above equation, the four equations of motion of the particle can be obtained as
\begin{equation}\label{6}
P_t=-[1-\frac{2M}{r}+\frac{e^{-r/\beta M}}{rM}(\frac{r^2}{\beta^2}+\frac{2Mr}{\beta}+2M^2)]\dot{t}=-E,
\end{equation}
\begin{equation}\label{7}
 P_r=[1-\frac{2M}{r}+\frac{e^{-r/\beta M}}{rM}(\frac{r^2}{\beta^2}+\frac{2Mr}{\beta}+2M^2)]^{-1}\dot{r},
\end{equation}
\begin{equation}\label{8}
P_\theta=r^2\dot{\theta}=0,
\end{equation}
\begin{equation}\label{9}
P_\phi=r^2\dot{\phi}=L.
\end{equation}
In the above equations, $E$ represents the energy per unit mass of the particle, and $L$ represents the orbital angular momentum per unit mass of the particle. From Eqs. \eqref{6} and \eqref{9}, it can be found that
\begin{equation}\label{10}
\dot{t} =\frac{E}{[1-\frac{2M}{r}+\frac{e^{-r/\beta M}}{rM}(\frac{r^2}{\beta^2}+\frac{2Mr}{\beta}+2M^2)]},
\end{equation}
\begin{equation}\label{11}
\dot{\phi}=\frac{L}{r^2}.
\end{equation}

\begin{figure*}[]
\includegraphics[width=1\textwidth]{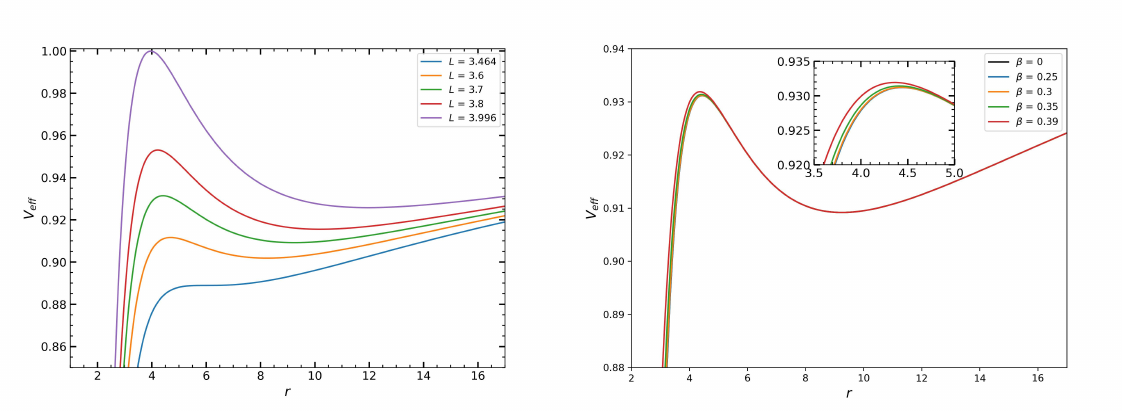}
\caption{The relationship diagram between the effective potential $V_{eff}$ and the radial radius $r$. Here, $M = 1$. In the first figure, fix $\beta = 0.35$. From the blue solid line to the purple solid line, $L$ is 3.464, 3.6, 3.7, 3.8, and 3.996 in turn. In the second figure, fix $L = 3.7$. From bottom to top, the solid lines of $\beta$ values are 0, 0.25, 0.3, 0.35, and 0.39 in turn.}
\label{b}
\end{figure*}

For a massive test particle, its four-dimensional velocity is a timelike unit vector. Therefore, there exists $g_{\mu\nu}{\dot{x}}^\mu{\dot{x}}^\nu=-1$. Substituting Eqs. \eqref{10} and \eqref{11} into equation \eqref{4} can simplify the Lagrangian to
\begin{equation}\label{12}
\begin{split}
&\frac{{\dot{r}}^2}{[1-\frac{2M}{r}+\frac{e^{-r/\beta M}}{rM}(\frac{r^2}{\beta^2}+\frac{2Mr}{\beta}+2M^2)]}+\frac{L^2}{r^2}\\&-\frac{E^2}{[1-\frac{2M}{r}+\frac{e^{-r/\beta M}}{rM}(\frac{r^2}{\beta^2}+\frac{2Mr}{\beta}+2M^2)]}=-1.
\end{split}
\end{equation}
Rewrite the above equation as
\begin{equation}\label{13}
{\dot{r}}^2=E^2-V_{eff},
\end{equation}
here
\begin{equation}\label{14}
V_{eff}=[1-\frac{2M}{r}+\frac{e^{-r/\beta M}}{rM}(\frac{r^2}{\beta^2}+\frac{2Mr}{\beta}+2M^2)](1+\frac{L^2}{r^2}).
\end{equation}
The above equation is the effective potential for the radial motion of the particle. We will conduct a detailed analysis of the properties of the effective potential in Fig.\ref{b}. From the first figure in Fig.\ref{b}, it can be obtained that as the orbital angular momentum decreases, the two extreme values of the effective potential gradually approach each other and finally converge at one point. This line corresponds to the innermost stable circular orbit (ISCO) of the particle's motion, that is, the blue solid line shown when $L = 3.464$. The maximum value of the effective potential corresponds to the marginally bound orbit (MBO) of the particle, that is, the purple solid line shown when $L = 3.996$. At this time, $V_{eff}=E=1$. These two kinds of orbits will be introduced in detail in the next section. From the second figure in Fig.\ref{b}, it can be obtained that as the hair parameter increases, the extreme value of the effective potential gradually increases. Here, when the hair parameter is equal to zero, it corresponds to the Schwarzschild black hole case. In addition, by combining Eqs. \eqref{13} and \eqref{14}, we find that when $r\rightarrow\infty$, $V = 1$. Therefore, when $E\geq1$, there is no bound orbit and the particle can move to infinity; when $E<1$, there is a bound orbit for the motion of the test particle. This is the focus of our attention, and the specific details will be presented in the next section.

\section{Bound orbits in the spacetime of a regular hairy black hole}\label{3.0}
In a strong gravitational field, particles around black holes will exhibit complex trajectories. Studying the properties of the trajectories of particles around black holes can indirectly obtain physical information about black holes. In this section, we primarily study the marginally bound orbit (MBO), the innermost stable circular orbit (ISCO), precession, and periodic orbits of the test particle.

\subsection{Relevant properties of MBO and ISCO}\label{3.1}
\begin{figure*}[]
\includegraphics[width=1\textwidth]{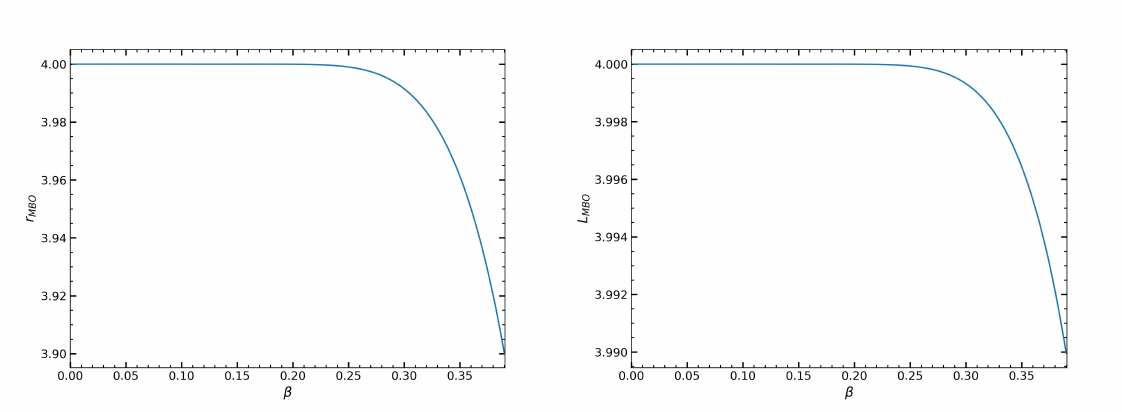}
\caption{The properties of the MBO of the particle around a regular hairy black hole. Take $M = 1$. The first figure is the $r_{MBO}-\beta$ relationship diagram; the second figure is the $L_{MBO}-\beta$ relationship diagram. Here $\beta$ is the hair parameter.}
\label{c}
\end{figure*}

The particle is in a marginally bound state on the MBO. A particle on this orbit has the same energy ($E = 1$) as a stationary particle at infinity. This orbit is not stable, and any slight additional perturbation will cause the particle to either escape to infinity from the black hole's gravitational field or spiral into the black hole. This orbit is determined by the following conditions
\begin{equation}\label{15}
V_{eff}=E=1,\ \ \partial_r{V_{eff}}=0.
\end{equation}
An analytical solution to the above equations is difficult to obtain. Therefore, we solve the equations numerically to plot the relationships between the radius of the MBO and the hair parameter $\beta$, as well as between $L_{MBO}$ and the hair parameter $\beta$, as shown in Fig.\ref{c}. It is obvious from Fig.\ref{c} that as the hair parameter $\beta$ increases, both $r_{MBO}$ and $L_{MBO}$ show a decreasing trend. In the initial stage, as the hair parameter $\beta$ changes, the change trends of $r_{MBO}$ and $L_{MBO}$ are not obvious. This indicates that at this time, the difference between a regular hairy black hole and a Schwarzschild black hole is very small and not easy to distinguish. When the hair parameter is close to the critical value $\beta_0$ of an extremal black hole, the difference between the corresponding orbital parameters and the Schwarzschild black hole is significantly enhanced. However, such differences are actually very small. In other words, even near the extremal black hole, the hair parameter only induces slight changes in the relevant parameters. These differences might be difficult to distinctly distinguish the characteristics of a hairy black hole from those of a Schwarzschild black hole in actual observations. This suggests that under certain conditions, the influence of the hair parameter becomes relatively weak.

\begin{figure*}[]
\includegraphics[width=1\textwidth]{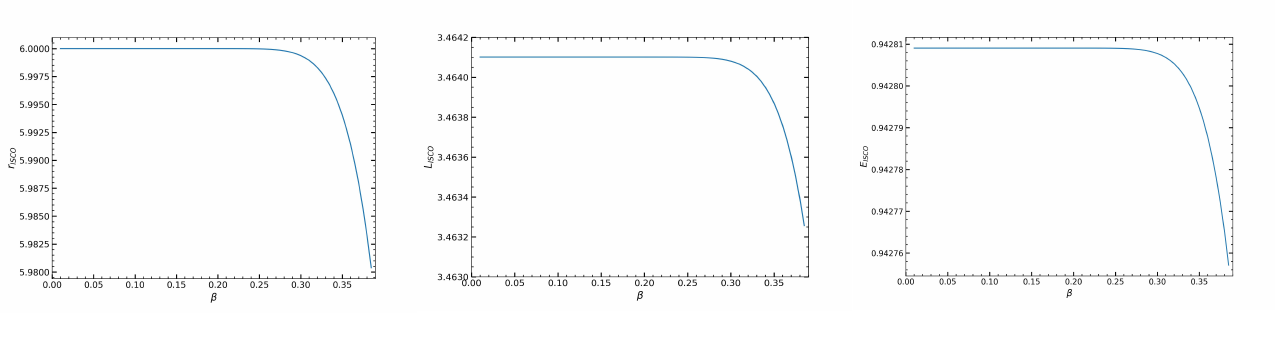}
\caption{The properties of the ISCO of the particle around a regular hairy black hole. Take $M = 1$. The first figure is the $r_{ISCO}-\beta$ relationship diagram; the second figure is the $L_{ISCO}-\beta$ relationship diagram; the third figure is the $E_{ISCO}-\beta$ relationship diagram.}
\label{d}
\end{figure*}

The ISCO is the orbit with the smallest radius at which a particle can maintain stable circular motion near a black hole. The ISCO is defined as
\begin{equation}\label{16}
\dot{r}=0,\ \ \partial_rV_{eff}=0,\ \ \partial_r^2V_{eff}=0.
\end{equation}
Since obtaining an analytical solution is challenging, we also use numerical methods here to plot the variations of the ISCO radius $r_{ISCO}$, the angular momentum $L_{ISCO}$, and the energy $E_{ISCO}$ with respect to the hair parameter $\beta$. As shown in Fig.\ref{d}, with the increase of the hair parameter $\beta$, the radius $r_{ISCO}$, angular momentum $L_{ISCO}$, and energy $E_{ISCO}$ of the ISCO all show a slow decreasing trend in the initial stage. Compared with the corresponding values of the Schwarzschild black hole, these changes are not significant. (In the figure, when the hair parameter $\beta = 0$, the regular hairy black hole degenerates into a Schwarzschild black hole. At this time, $r_{ISCO}= 6$, which is consistent with the calculation result of the Schwarzschild black hole\cite{Gao:2020wjz,Pugliese:2011py}.) However, as the parameter $\beta$ approaches the extremal black hole regime, although $r_{ISCO}$, $L_{ISCO}$, and $E_{ISCO}$ exhibit certain changes in theoretical calculations—and these changes gradually intensify as $\beta$ approaches the critical value $\beta_0$—the differences in these orbital parameters are difficult to distinctly differentiate the characteristics of a regular hairy black hole from those of a Schwarzschild black hole in actual observations. Particularly near the critical value, while these parameter changes are theoretically quantifiable, their magnitude is extremely small, making it challenging to exceed the sensitivity limits of current observational techniques.

\begin{figure}[]
\includegraphics[width=0.5\textwidth]{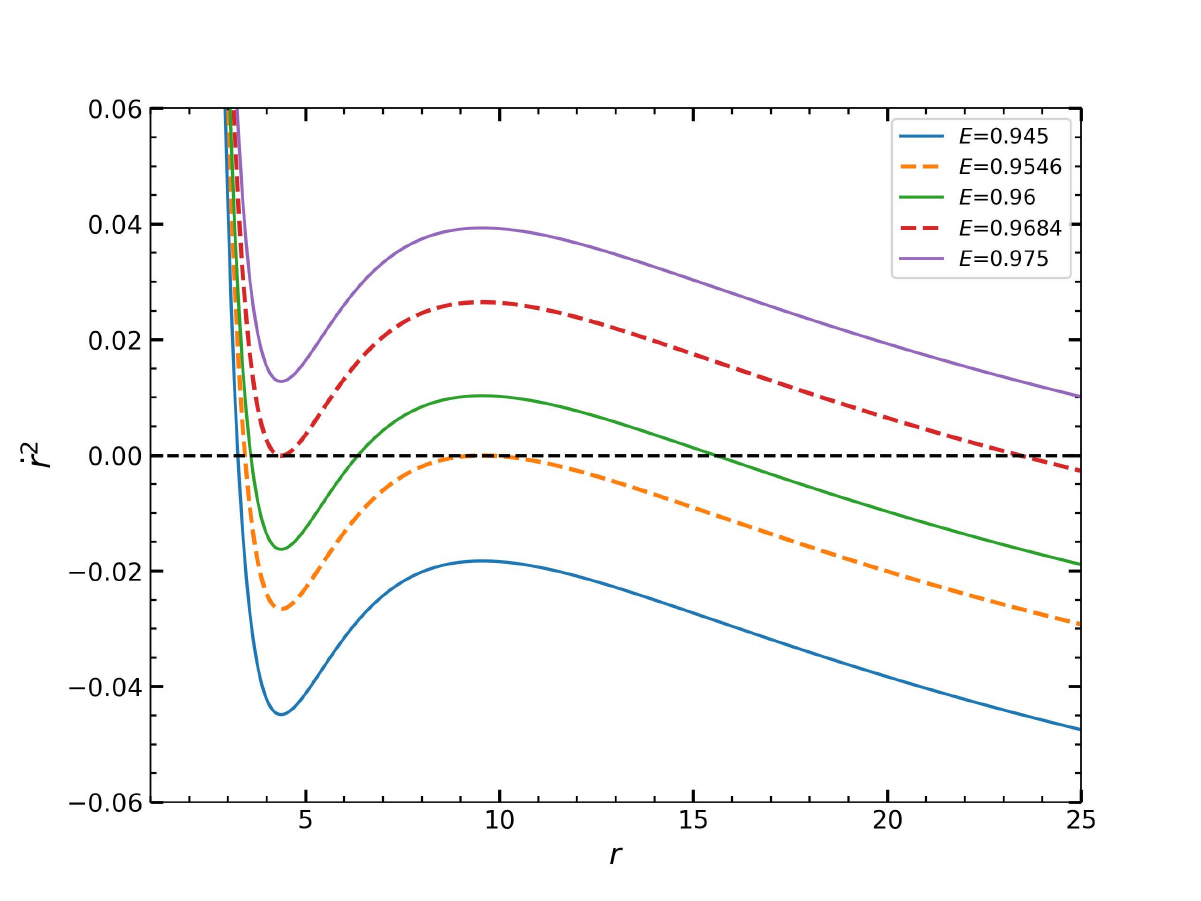}
\caption{The relationship diagram of ${\dot{r}}^2 - r$. Here, take $M = 1$, $\beta = 0.25$, and $L=0.5(L_{MBO}+L_{ISCO})$.}
\label{e}
\end{figure}

\begin{figure}[]
\includegraphics[width=0.5\textwidth]{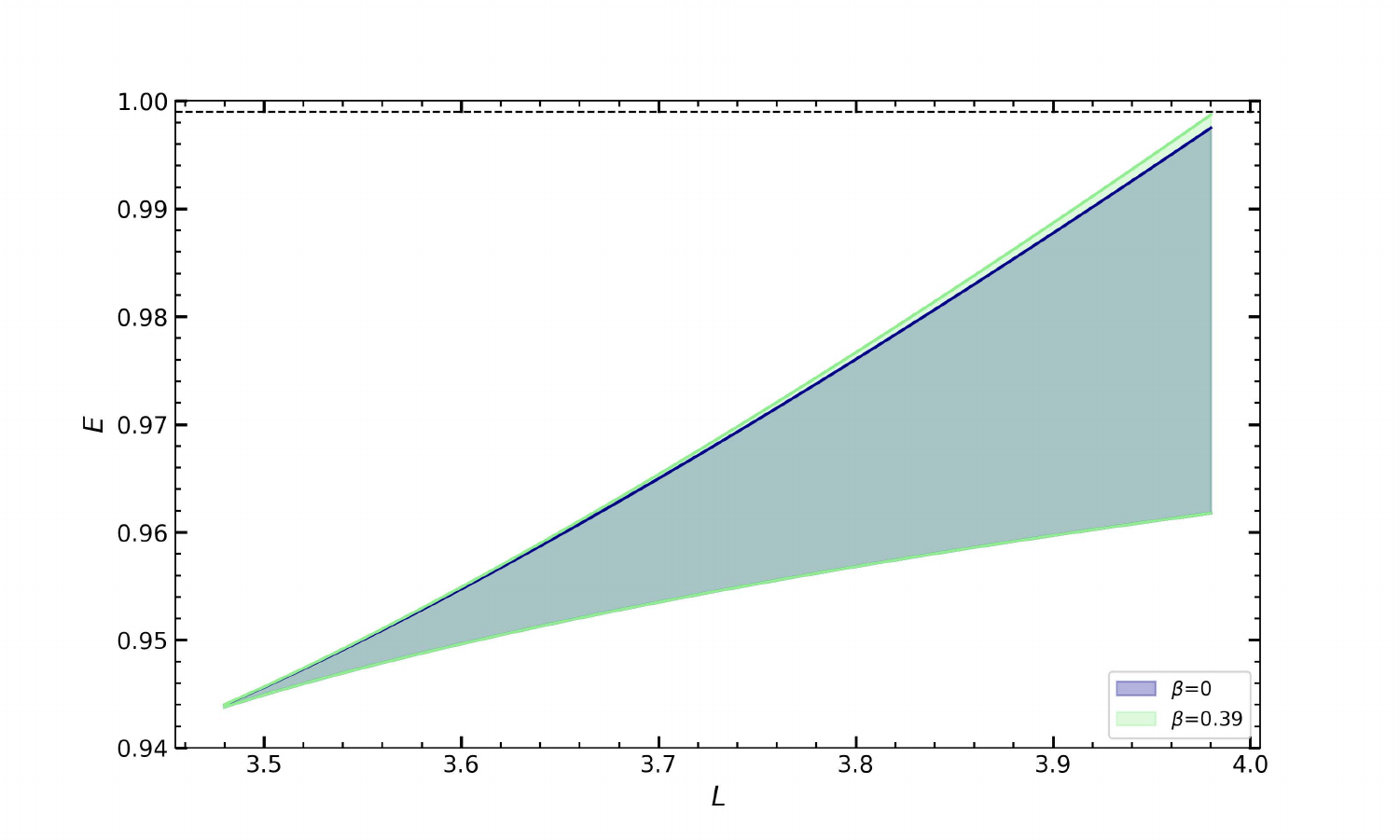}
\caption{The allowed parameter space of the bound orbit $(L, E)$ of the test particle around the regular hairy black hole. Here, M = 1.}
\label{f}
\end{figure}

In addition, according to the previous analysis, we find that there should be an allowed region for the orbital angular momentum and energy of the particle that satisfy the bound orbit motion. As shown in Fig.\ref{e}, when $\beta = 0.25$ and $L=0.5(L_{MBO}+L_{ISCO})$, the energy $E$ of the bound orbit is in the range of (0.9546, 0.9684). Secondly, it can also be obtained from Fig.\ref{e} that when the two extreme values of the ${\dot{r}}^2 - r$ curve are of opposite signs, the orbit represented at this time is a bound orbit. Further, the $(L, E)$ region of the bound orbit can be obtained, as shown by the shaded area in Fig.\ref{f}. In Fig.\ref{f}, we have plotted the $(L, E)$ parameter space for $\beta = 0$ and $\beta = 0.39$. Obviously, as the parameter $\beta$ increases, the parameter space has a tendency to decrease and the allowed energy range of the bound orbit of the test particle increases. The uppermost and lowermost boundary points of the parameter space are $E_{MBO}$ and $E_{ISCO}$ corresponding to different orbital angular momenta respectively. 

\subsection{Precession and periodic orbits}\label{3.2}
Precession and periodic orbits are important tools for studying the dynamical properties of test particles around black holes. Since the regular hairy black hole we are investigating is static and spherically symmetric, the motion of a particle is described only by the radial coordinate $r$ and the angular coordinate $\phi$. As shown in Fig.\ref{e}, bound orbits have two turning points, $r_1$ and $r_2$, and the particle's motion oscillates back and forth between these two turning points. During a complete oscillation cycle, the angular displacement of the particle is given by
\begin{equation}\label{17}
\Delta\phi=\oint d\phi=2\int_{r_1}^{r_2}\frac{d\phi}{dr}dr.
\end{equation}
In this study, we adopt the classification method for bound orbits proposed in \cite{Levin:2008mq}, where these orbits are described by a unique parameter $q$. The parameter $q$ is defined as
\begin{equation}\label{18}
q=\frac{\Delta\phi}{2\pi}-1,
\end{equation}
evidently, from the equation above, when $q$ is an irrational number, the test particle does not return to its original position after completing one orbit around the black hole, resulting in a deviation. These orbits exhibit orbital precession with each cycle\cite{2018grav}. This precession effect manifests as the particle's trajectory gradually deviating from its initial state, with each cycle producing a precession angle $\Delta\omega = \Delta\phi - 2\pi$, which represents the offset angle relative to the previous cycle. When $q$ is a rational number, the particle's trajectory exhibits periodicity. Such periodic orbits mean that the particle can return to its original position within a finite amount of time, forming a completely closed trajectory and thus presenting a periodic structure. As stated in \cite{Levin:2008mq}, quasi-periodic orbits (i.e., precessing orbits) can be obtained by perturbing periodic orbits. Therefore, an in-depth study of periodic orbits can not only reveal the orbital dynamics of the particle around a regular hair black hole but also provide important theoretical insights for exploring the spacetime structure surrounding black holes. In addition, the analysis of periodic orbits can offer new perspectives for understanding the trajectory evolution of particles in the gravitational field of black holes and potential radiation mechanisms.

For precessing orbits, the major axis of the particle's orbit slowly rotates in space, causing the subsequent orbit to deviate from the previous one and resulting in an orbital precession angle $\Delta\omega$. This precession effect, induced by the central compact object, is also known as Schwarzschild precession. Based on the observational results of the S2 star orbiting the supermassive black hole Sgr A* as presented in \cite{GRAVITY:2020gka}, we will attempt to use these observational data to constrain the regular hairy black hole. 

According to \cite{GRAVITY:2020gka}, the measured Schwarzschild precession of the S2 star, $\Delta\omega_{S2}$, is related to the Schwarzschild precession predicted by general relativity, $\Delta\omega_{GR}$, by the ratio
\begin{equation}\label{19}
f_{sp}=\frac{\Delta\omega_{S2}}{\Delta\omega_{GR}}=1.10\pm0.19,
\end{equation}
evidently, this ratio exhibits significant uncertainty, indicating a certain deviation between the actual observational results and the theoretical predictions of general relativity. This deviation provides an opportunity to test the applicability of general relativity and to constrain the parameters of alternative theories. By comparing alternative theories (such as the regular hairy black hole model) with observational data, the theoretical parameters can be further adjusted to better match the actual observations.

\begin{figure}[]
\includegraphics[width=0.45\textwidth]{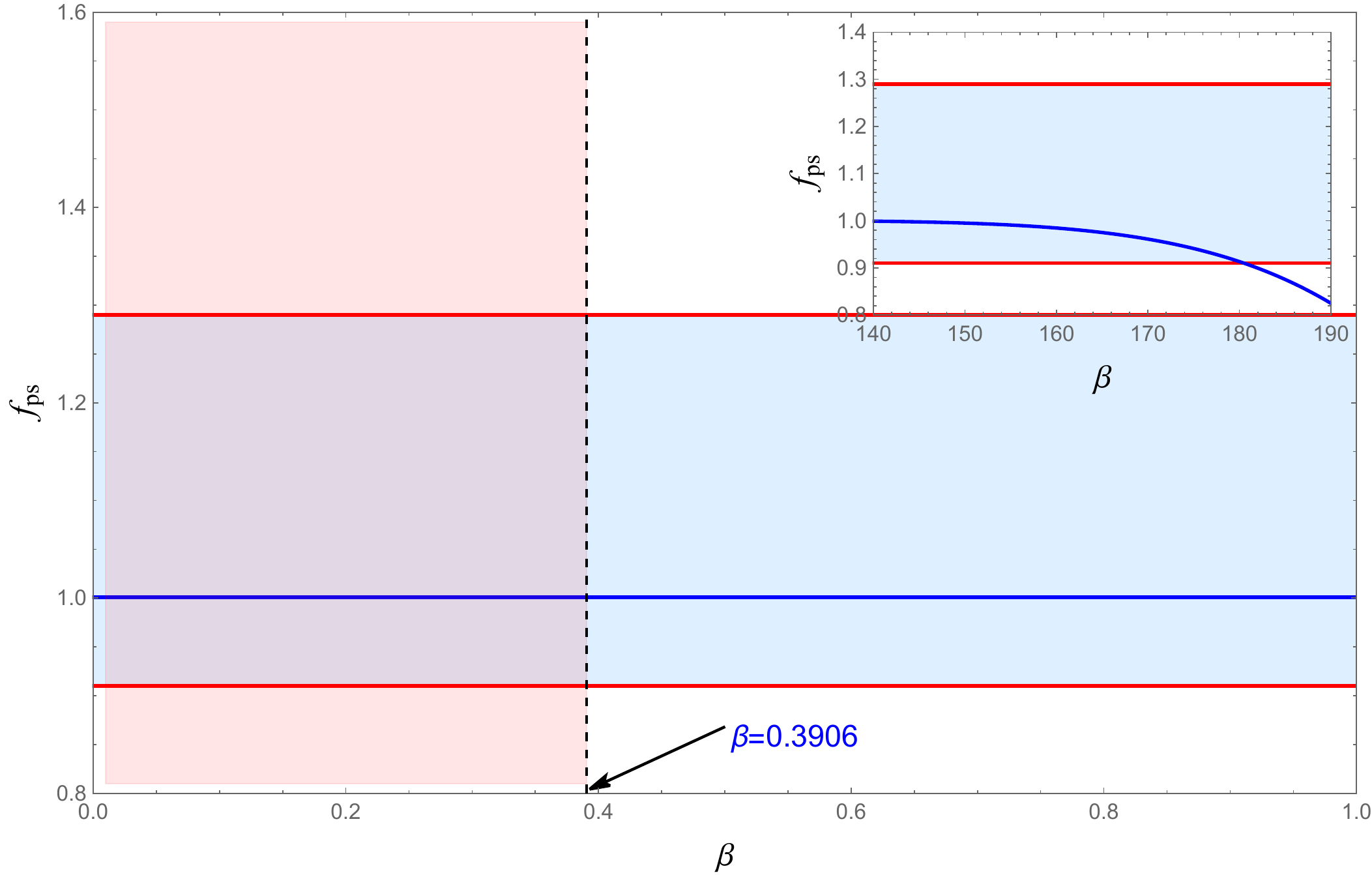}
\caption{The precession data of the S2 star is used to constrain the hair parameter $\beta$.}
\label{aa}
\end{figure}

If the regular hairy black hole is assumed to be a candidate model for the supermassive black hole Sgr A* at the center of the Milky Way, the motion trajectories of the particle around this black hole can be described as\cite{Bambi:2012ku}
\begin{equation}\label{20}
r=\frac{a(1-e^2)}{1+e\cos{\Psi}},
\end{equation}
here, $a$ and $e$ represent the semi-major axis and eccentricity of the orbit, respectively, $r_a = a(1-e)$ is the periapsis, $r_b = a(1+e)$ is the apoapsis, and $\Psi$ is the angle between the orbital semi-major axis and the radial direction of the orbit. Therefore, during orbital evolution, the angular displacement \eqref{17} can be rewritten as
\begin{equation}\label{21}
\begin{split}
\Delta\phi&=2\int_{r_a}^{r_b}\frac{d\phi}{dr}dr=2\int_{0}^{\pi}\frac{d\phi}{dr}\frac{dr}{d\Psi}d\Psi\\&=2\int_{0}^{\pi}\frac{ae\left(1-e^2\right)L\sin{\Psi}}{r^2{(1+e\cos{\Psi)}}^2\sqrt{E^2-f(r)(1+\frac{L^2}{r^2})}}d\Psi,
\end{split}
\end{equation}
here, $E$ and $L$ are
\begin{equation}\label{22}
E^2=\frac{f(r_a)f(r_b)(r_a^2-r_b^2)}{r_a^2f\left(r_b\right)-r_b^2f(r_a)},
\end{equation}
\begin{equation}\label{23}
L^2=\frac{r_a^2r_b^2(f\left(r_a\right)-f\left(r_b\right))}{r_a^2f\left(r_b\right)-r_b^2f(r_a)}.
\end{equation}
Thus, the precession angle of the regular hairy black hole, $\Delta\omega_R$, can be expressed as
\begin{equation}\label{24}
\begin{split}
\Delta\omega_R&\approx\Delta\phi-2\pi\\&=2\int_{0}^{\pi}\frac{ae\left(1-e^2\right)L\sin{\Psi}}{r^2{(1+e\cos{\Psi)}}^2\sqrt{E^2-f(r)(1+\frac{L^2}{r^2})}}d\Psi-2\pi.
\end{split}
\end{equation}
The precession of the particle around a regular hairy black hole can be derived from equation \eqref{19} as
\begin{equation}\label{25}
f_{sp}=\frac{\Delta\omega_R}{\Delta\omega_{GR}},
\end{equation}
here
\begin{equation}\label{26}
\Delta\omega_{GR}=\frac{6\pi M}{a(1-e^2)}.
\end{equation}
Due to the complex exponential terms in the metric function $f(r)$, analytical solutions are difficult to obtain. Therefore, we use numerical methods for the calculations. As shown in Fig.\ref{aa}, in the presence of a black hole ($0 \leq \beta \leq 0.3906$), the correction effect of the hair parameter is consistent with the current observational constraints. However, when the parameter value approaches 180, it exceeds the observational constraints, but at this point, the spacetime no longer contains a black hole. This indicates that the influence of the hair parameter on precessing orbits is very weak, making it difficult to impose strong constraints on it through precessing orbits. 

For periodic orbits, the test particle follows a bound orbit that returns to its initial state after a finite period of motion. According to the classification method of particle periodic orbits proposed in the literature\cite{Levin:2008mq}, use three integers $(z, w, v)$ to define a rational number $q$, and establish the relationship between it and a definite periodic orbit. Here, $q$ is defined by the following formula.
\begin{equation}\label{27}
q=\frac{\Delta\phi}{2\pi}-1=w+\frac{v}{z},
\end{equation}
among them, $\Delta\phi$ is the accumulated azimuth angle between consecutive apoapsis in a periodic orbit, $z$ is the number of leaves of the orbit, $w$ is the number of rotations around the center, and $v$ is the number of leaves skipped by the orbit when reaching the apex. Here
\begin{equation}\label{28}
\Delta\phi=\oint d\phi=2\int_{r_1}^{r_2}\frac{d\phi}{dr}dr.
\end{equation}
Combining Eqs.\eqref{10} and \eqref{11}, equation \eqref{27} will be rewritten as
\begin{widetext}
\begin{equation}\label{29}
q=\frac{1}{\pi}\int_{r_1}^{r_2}\frac{\dot{\phi}}{\dot{r}}dr-1=\frac{1}{\pi}\int_{r_1}^{r_2}\frac{L}{r^2 \sqrt{E^2-[1-\frac{2M}{r}+\frac{e^{-r/\beta M}}{rM}(\frac{r^2}{\beta^2}+\frac{2Mr}{\beta}+2M^2)](1+\frac{L^2}{r^2})} }dr-1.
\end{equation}
\end{widetext}
Here, $r_1$ and $r_2$ are the periapsis radius and the apoapsis radius.

\begin{figure*}[htbp]
    \centering
    
    \begin{minipage}[t]{0.5\textwidth}
        \centering
        \includegraphics[width=\textwidth]{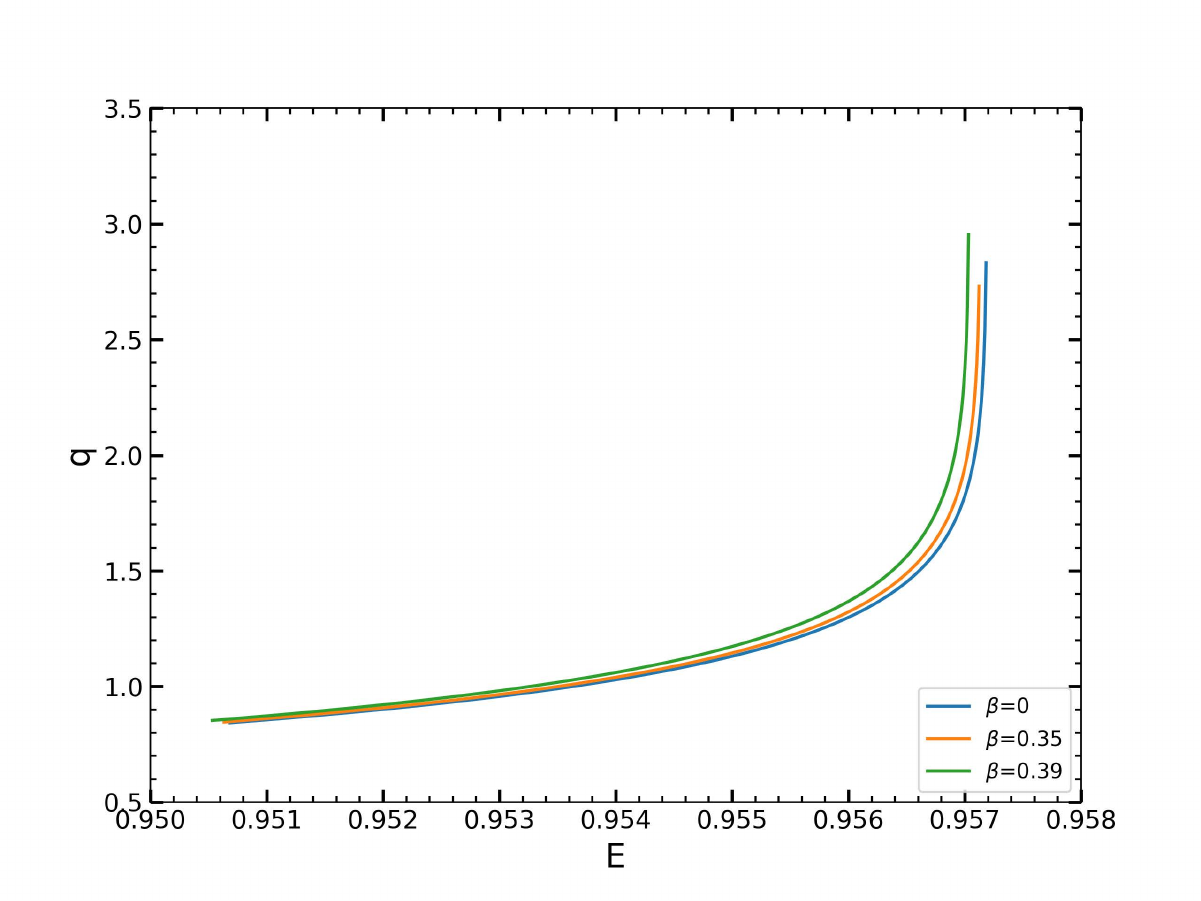} 
        \captionsetup{labelformat=empty} 
        \caption*{(a)$L=0.3L_{MBO}+0.7L_{ISCO}$} 
    \end{minipage}%
    \hfill
    \begin{minipage}[t]{0.5\textwidth}
        \centering
        \includegraphics[width=\textwidth]{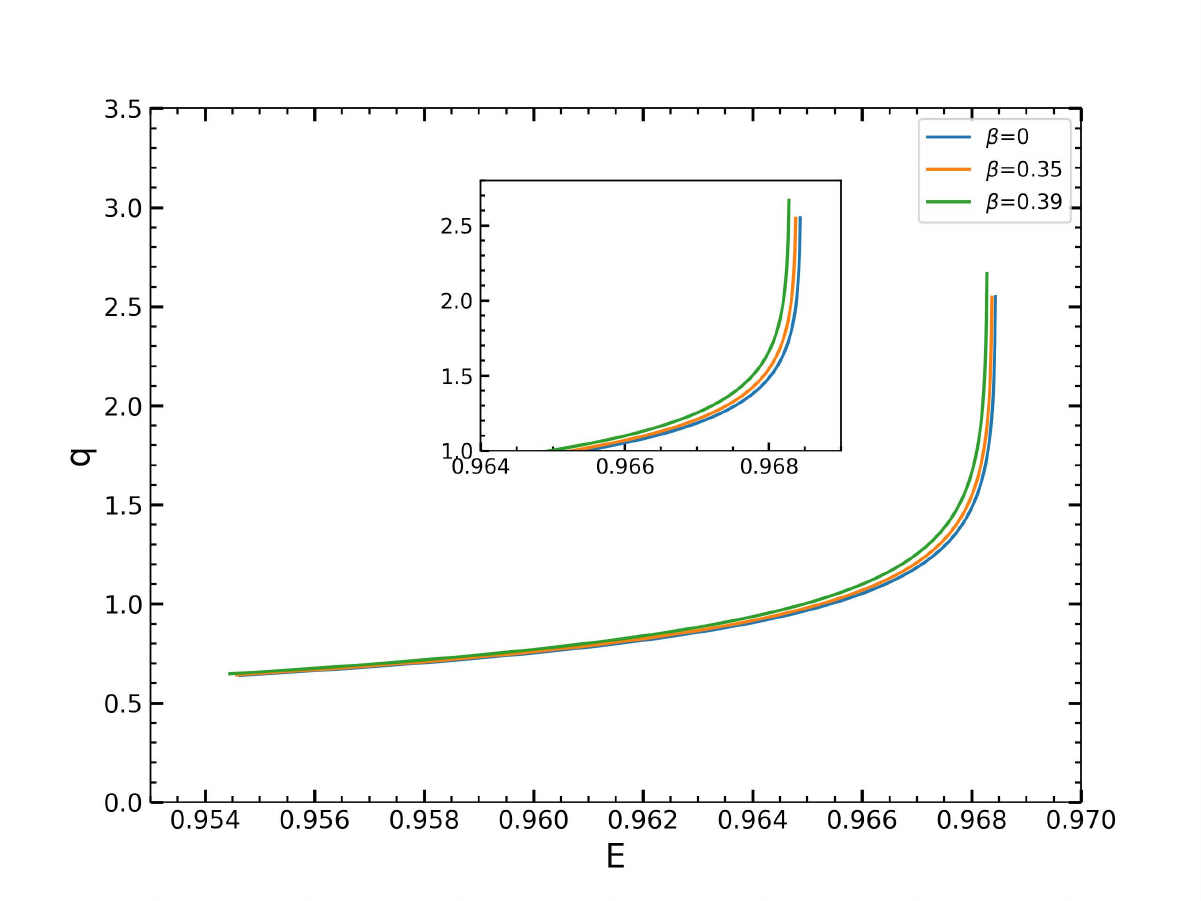} 
        \captionsetup{labelformat=empty} 
        \caption*{(b)$L=0.5{(L}_{MBO}+L_{ISCO})$} 
    \end{minipage}
     
    \begin{minipage}[t]{0.5\textwidth}
        \centering
        \includegraphics[width=\textwidth]{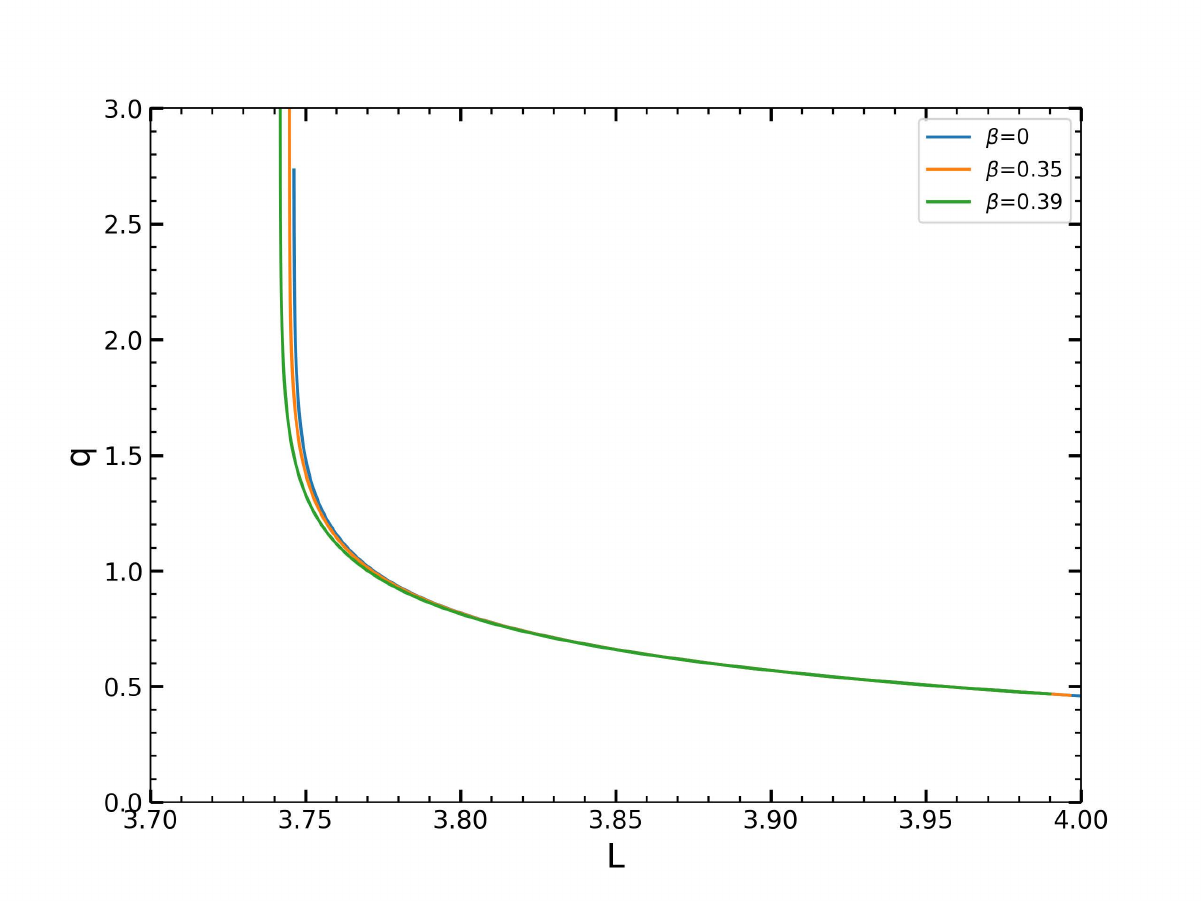} 
        \captionsetup{labelformat=empty}
        \caption*{(c)$E=0.97$}
    \end{minipage}%
    \hfill
    \begin{minipage}[t]{0.5\textwidth}
        \centering
        \includegraphics[width=\textwidth]{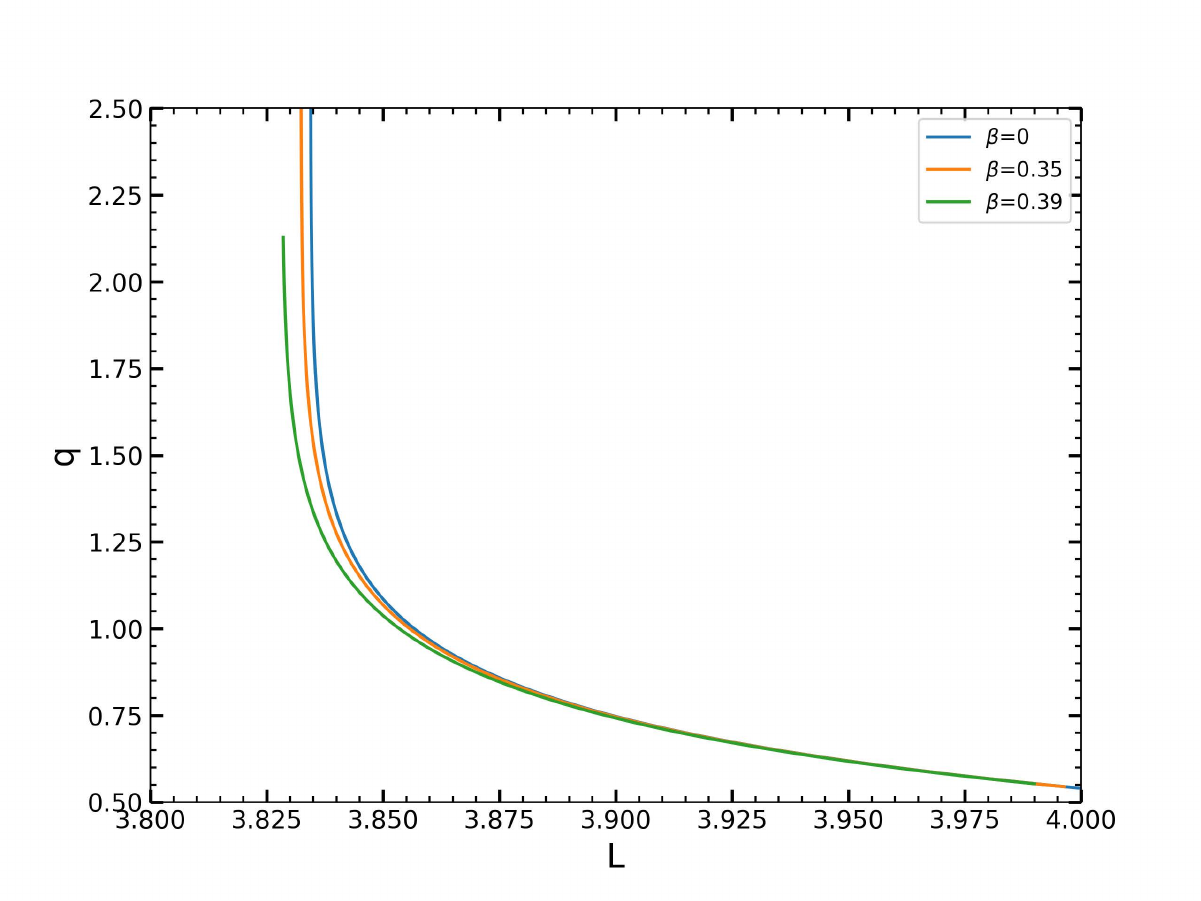}
        \captionsetup{labelformat=empty}
        \caption*{(d)$E=0.98$}
    \end{minipage}
    
    \caption{Figures (a) and (b) are the relationship diagrams of the rational number $q$ and energy $E$ corresponding to different orbital angular momenta $L$ and different hair parameters $\beta$; Figures (c) and (d) are the relationship diagrams of the rational number $q$ and orbital angular momentum $L$ corresponding to different energies $E$ and different hair parameters $\beta$. Among them, $M = 1$, $\beta = 0, 0.35, 0.39$.} 
    \label{g}
\end{figure*}

As known from Fig.\ref{f}, there is an allowed parameter space for energy $E$ and orbital angular momentum $L$. Therefore, according to equation \eqref{29}, by fixing one of $E$ or $L$ and the parameter $\beta$, the relationship diagram of $q$ and $E$ or $L$ can be drawn. As shown in Fig.\ref{g}. From figures (a) and (b) of Fig.\ref{g}, it can be obtained that different orbital angular momenta correspond to different energy value ranges, which is in line with the situation shown in Fig.\ref{f}. From the $q-E$ relationship diagram, it can also be analyzed that as the energy $E$ increases, $q$ increases slowly. When the energy $E$ is in the extreme value situation, the rational number $q$ increases sharply. And the extreme value of energy $E$ decreases as the parameter $\beta$ increases. From figures (c) and (d) of Fig.\ref{g}, it can be obtained that the rational number $q$ decreases as the orbital angular momentum $L$ increases, and there is a sharp decrease in the rational number $q$ at the extreme value of the orbital angular momentum. Moreover, the extreme value of the orbital angular momentum $L$ decreases as the parameter $\beta$ increases.

In addition, we performed numerical calculations of the energy and orbital angular momentum for different periodic orbits ($z, w, v$) under fixed orbital angular momentum $L$ or energy $E$ and parameter $\beta$, as shown in Tables \ref{table1} and \ref{table2}. 
The results show that for the same periodic orbit, as the parameter $\beta$ increases, both the orbital energy and angular momentum decrease. However, the magnitude of this change is relatively small, making it difficult to effectively distinguish between the regular hairy black hole and the Schwarzschild black hole through observations of periodic orbit dynamics. This indicates that the modification effect of the regular hairy black hole model on orbital dynamics is weak in terms of energy and angular momentum, further revealing the parameter degeneracy between the two black hole models under current observational conditions. Nevertheless, these subtle differences provide a theoretical foundation for exploring the characteristics of the regular hairy black hole through higher-precision observational methods in the future.

\begin{table*}[]
\centering
\begin{tabular}{p{3.2cm}p{0.8cm}p{1.7cm}p{1.7cm}p{1.7cm}p{1.7cm}p{1.7cm}p{1.7cm}p{1.7cm}p{1.7cm}}
\hline\hline
\rule{0pt}{12pt}$L$ & $\beta$ & $E_{(1,1,0)}$ & $E_{(1,2,0)}$ & $E_{(2,1,1)}$ & $E_{(2,2,1)}$ & $E_{(3,1,2)}$ & $E_{(3,2,2)}$ & $E_{(4,1,3)}$ & $E_{(4,2,3)}$ \\
\hline
\rule{0pt}{12pt} 
$0.3L_{MBO}+0.7L_{ISCO}$ & 0 & 0.95362822 & 0.95708623 & 0.95660722 & 0.95717015 & 0.95686429 & 0.95717549 & 0.95694584 & 0.95718075 \\
\rule{0pt}{12pt} 
& 0.35 & 0.95350389 & 0.95702204 & 0.95652338 & 0.95711202 & 0.95678934 & 0.95712081 & 0.95687439 & 0.95712368 \\
\rule{0pt}{12pt} 
& 0.39 & 0.95326229 & 0.95691210 & 0.95637458 & 0.95701362 & 0.95665853 & 0.95702391 & 0.95675041 & 0.95702732 \\
\rule{0pt}{12pt} 
$0.5(L_{MBO}+L_{ISCO})$ & 0 & 0.96542534 & 0.96838291 & 0.96802649 & 0.96843431 & 0.96822485 & 0.96843848 & 0.96828496 & 0.96843977 \\
\rule{0pt}{12pt} 
& 0.35 & 0.96525130 & 0.96831201 & 0.96792931 & 0.96836995 & 0.96814048 & 0.96837483 & 0.96820548 & 0.96837635 \\
\rule{0pt}{12pt} 
& 0.39 & 0.96494576 & 0.96820269 & 0.96777272 & 0.96827210 & 0.96800689 & 0.96827825 & 0.96807995 & 0.96828020 \\
\hline\hline
\end{tabular}
\caption{Energy values corresponding to different periodic orbits $(z, w, v)$. Here, take $L = 0.3L_{MBO}+0.7L_{ISCO}$, $L = 0.5{(L}_{MBO}+L_{ISCO})$ and $M = 1$, and the parameter $\beta = 0, 0.35, 0.39$.}
\label{table1}
\end{table*}

\begin{table*}[]
\centering
\begin{tabular}{p{1.2cm}p{1.2cm}p{1.8cm}p{1.8cm}p{1.8cm}p{1.8cm}p{1.8cm}p{1.8cm}p{1.8cm}p{1.8cm}}
\hline\hline
\rule{0pt}{12pt}$E$ & $\beta$ & $L_{(1,1,0)}$ & $L_{(1,2,0)}$ & $L_{(2,1,1)}$ & $L_{(2,2,1)}$ & $L_{(3,1,2)}$ & $L_{(3,2,2)}$ & $L_{(4,1,3)}$ & $L_{(4,2,3)}$ \\
\hline
\rule{0pt}{12pt} 
0.97 & 0 & 3.77197612 & 3.74673495 & 3.74981831 & 3.74629399 & 3.74810031 & 3.74625875 & 3.74757974 & 3.74624803 \\
\rule{0pt}{12pt} 
& 0.35 & 3.77144205 & 3.74538012 & 3.74868056 & 3.74488566 & 3.74685797 & 3.74484460 & 3.74629938 & 3.74483195 \\
\rule{0pt}{12pt} 
& 0.39 & 3.77010706 & 3.74254238 & 3.74621973 & 3.74195436 & 3.74421508 & 3.74190288 & 3.74359045 & 3.74188674 \\
\rule{0pt}{12pt} 
0.98 & 0 & 3.85671457 & 3.83477572 & 3.83720996 & 3.83446575 & 3.83582295 & 3.83444322 & 3.83541529 & 3.83443659 \\
\rule{0pt}{12pt} 
& 0.35 & 3.85573037 & 3.83275139 & 3.83542644 & 3.83238990 & 3.83391929 & 3.83236232 & 3.83346947 & 3.83235406 \\
\rule{0pt}{12pt} 
& 0.39 & 3.85351356 & 3.82875540 & 3.83183053 & 3.82830572 & 3.83012430 & 3.82826912 & 3.82960461 & 3.82825792 \\
\hline\hline
\end{tabular}
\caption{Orbital angular momentum values corresponding to different periodic orbits $(z, w, v)$. Here, take $E = 0.97$, $E = 0.98$ and $M = 1$, and the parameter $\beta = 0, 0.35, 0.39$.}
\label{table2}
\end{table*}

\begin{figure*}[]
\includegraphics[width=1\textwidth]{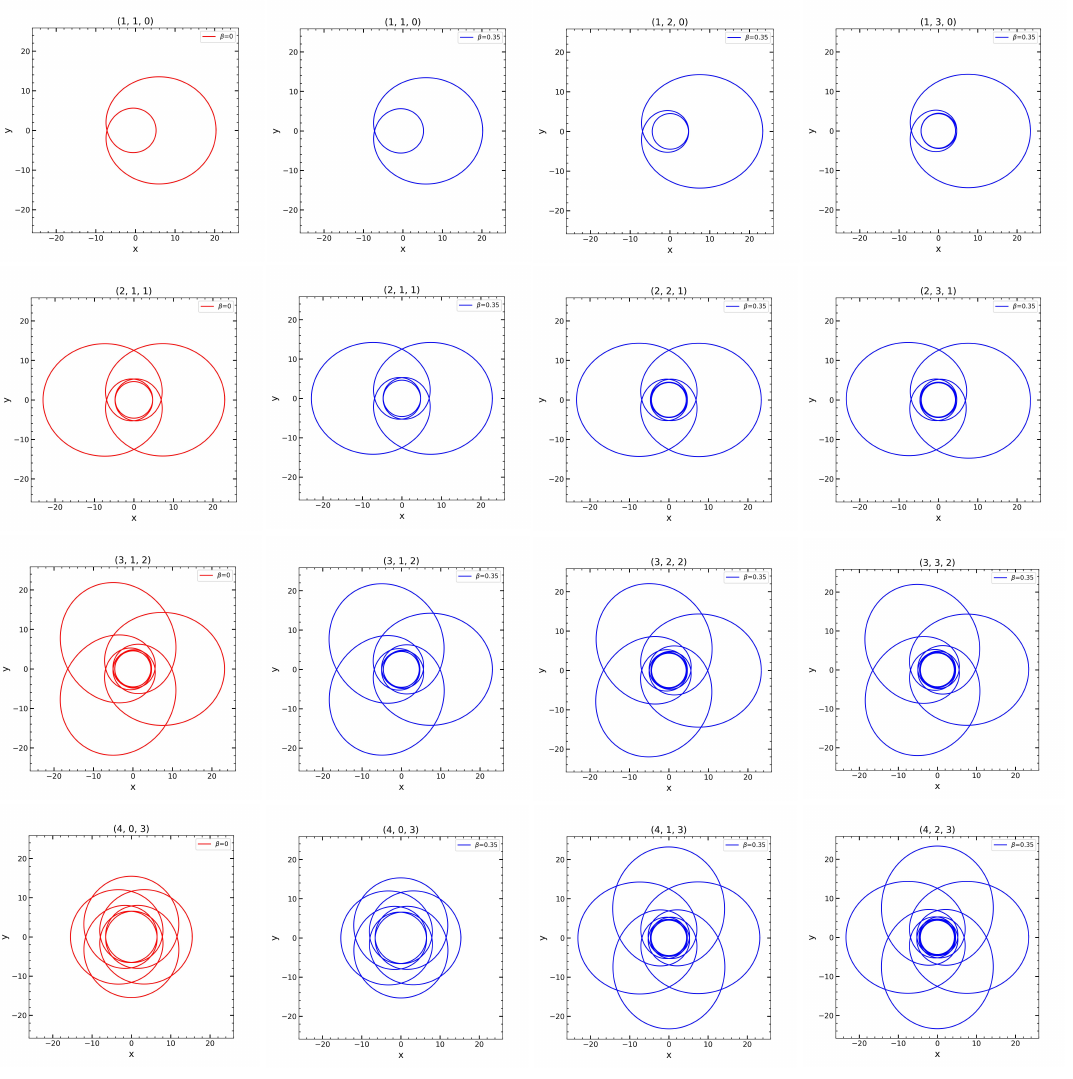}
\caption{Periodic trajectory diagrams of different periodic orbits $(z, w, v)$. Among them, the red solid line is the periodic trajectory of the Schwarzschild black hole; the blue solid line is the periodic trajectory of the regular hairy black hole with parameter $\beta = 0.35$. Here, the orbital angular momentum $L=0.5(L_{MBO}+L_{ISCO})$ and $M = 1$ are fixed.}
\label{h}
\end{figure*}

\begin{figure*}[]
\includegraphics[width=1\textwidth]{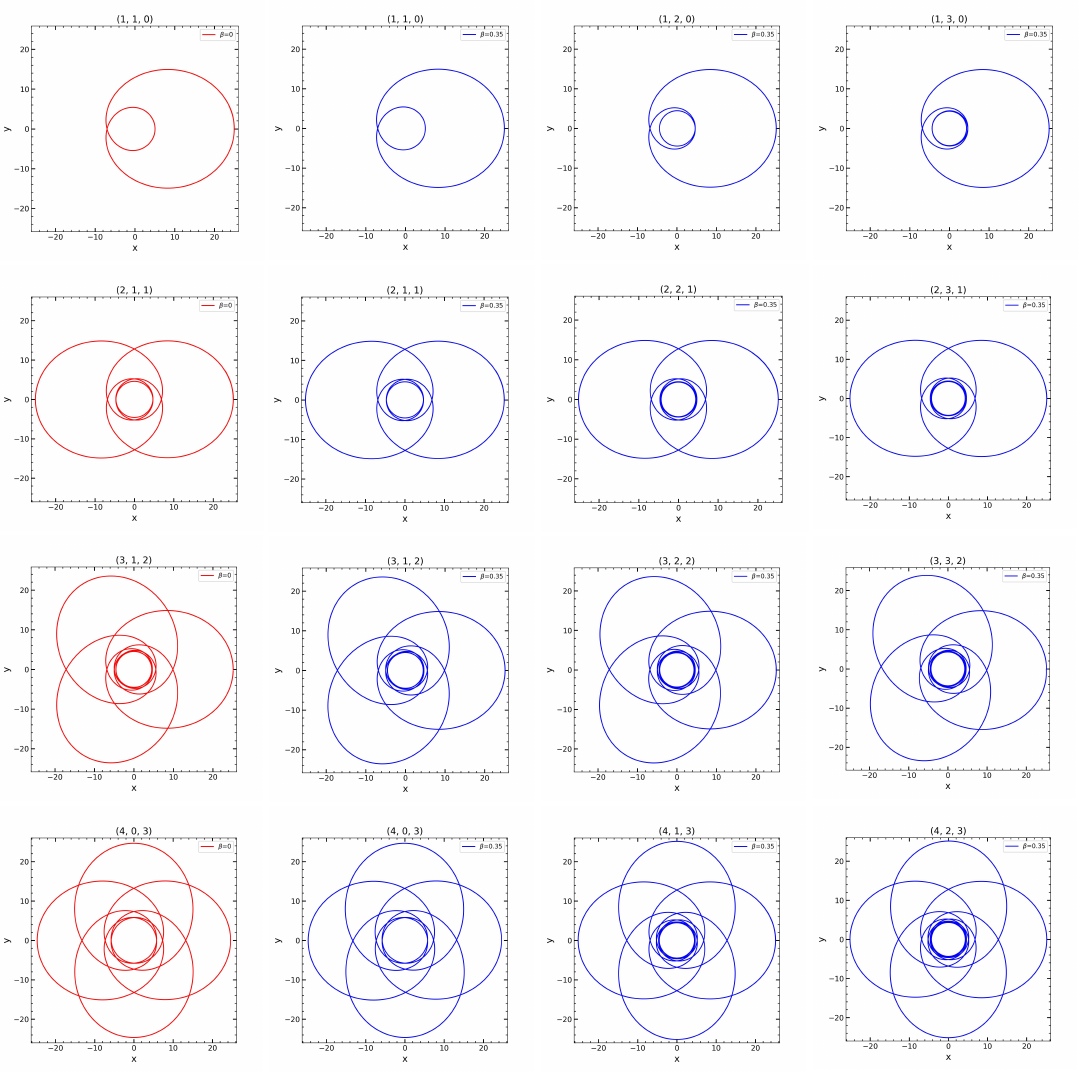}
\caption{Periodic trajectory diagrams of different periodic orbits $(z, w, v)$. Among them, the red solid line is the periodic trajectory of the Schwarzschild black hole; the blue solid line is the periodic trajectory of the regular hairy black hole with parameter $\beta = 0.35$. Here, the energy $E = 0.97$ and $M = 1$ are fixed.}
\label{i}
\end{figure*}

Next, with the relevant parameters fixed, we have drawn the periodic trajectory diagrams with different periodic orbits $(z, w, v)$ in Figs.\ref{h} and \ref{i}. In Fig.\ref{h}, with the orbital angular momentum $L=0.5(L_{MBO}+L_{ISCO})$ fixed, the periodic trajectories of different rational numbers $q$ for the Schwarzschild black hole with $\beta = 0$ are drawn respectively, as shown by the red solid line in Fig.\ref{h}; the periodic trajectory of the regular hairy black hole with $\beta = 0.35$ is shown by the blue solid line in Fig.\ref{h}. In Fig.\ref{i}, with the energy $E = 0.97$ fixed, the periodic trajectories of the Schwarzschild black hole (red solid line) and the periodic trajectory of the regular hairy black hole with $\beta = 0.35$ (blue solid line) are also drawn. Combining Figs.\ref{h} and \ref{i}, it is clearly found that as $z$ and $w$ increase, the number of leaves of the periodic trajectory and the linear density number of rotations around the center increase, that is, the trajectory appears more complex. 

Additionally, for the same periodic orbit, compared to the Schwarzschild black hole, the periapsis radius of the test particle in a regular hairy black hole is slightly closer to the black hole, while the apoapsis radius is slightly farther from the black hole. Although these differences are numerically small (see Tables \ref{table1} and \ref{table2}), they reflect the slight modifications to the spacetime geometry of the black hole introduced by the hair parameter $\beta$. Specifically, this modification allows the test particle to enter stable periodic orbits with lower energy or angular momentum, thereby revealing the unique dynamical characteristics of the regular hairy black hole.

Although these effects are relatively weak, further analysis of the dynamical characteristics of these periodic orbits can reveal the unique gravitational wave radiation features of the regular hairy black hole, particularly their impact on gravitational wave signals in the extreme-mass-ratio inspiral (EMRI) system. In EMRI systems, periodic orbits play an important role as transitional orbits and are accompanied by gravitational wave radiation\cite{Levin:2008mq,Glampedakis:2002ya}. This characteristic motivates us to preliminarily explore the periodic orbits of timelike test particle around the regular hairy black hole and their associated gravitational wave signals. The relevant content will be discussed in detail in the next section to further deepen our understanding of the dynamical properties of a regular hairy black hole and their connection to gravitational wave astronomy. This study also aims to provide potential research pathways for distinguishing different black hole models through precise gravitational wave observations in the future.

\section{Gravitational wave radiation of periodic orbits}\label{4.0}
This section will study the gravitational wave radiation characteristics of periodic orbits of the particle around a regular hairy black hole. In the extreme-mass-ratio inspiral (EMRI) system, the periodic motion of timelike test particle emits gravitational waves, gradually reducing the orbital angular momentum and energy. This causes the particle's orbit to slowly spiral closer to the black hole, eventually merging with the supermassive black hole. Since this process usually lasts for a long time (about $10^3$ to $10^5$ orbital periods), the reduction in orbital angular momentum and energy during a single orbit is very small compared to the total energy of the system. Therefore, it is reasonable to use the adiabatic approximation condition\cite{Hughes:2005qb,Sundararajan:2007jg} for the study. Given that our goal is to evaluate whether the hair parameter can be detected through gravitational wave signals and that the number of orbital periods is extremely large, the adiabatic approximation is reasonable and sufficient in this scenario. In addition, to quickly and efficiently obtain the gravitational wave radiation waveforms of the particle on periodic orbits, we adopted the Kludge waveform method \cite{Babak:2006uv}. This method ensures sufficient accuracy while maintaining computational efficiency, providing reliable tool support for exploring the gravitational wave characteristics of the regular hairy black hole.

The generation of Kludge waveforms is divided into two stages: (1) orbital evolution (the periodic orbits discussed in detail in the previous section); (2) gravitational wave waveforms can be constructed for any periodic orbit. In the second stage, using the quadrupole formula for gravitational radiation\cite{Thorne:1980ru,Liang:2022gdk}, the gravitational wave waveforms emitted by the orbit can be calculated. To ensure consistency in the units of the formula, the gravitational constant $G$ and the speed of light $c$ are retained in the calculations. Specifically, the quadrupole formula can be expressed as 
\begin{equation}\label{30}
h_{ij}=\frac{4G\eta M}{c^4D_L}(\upsilon_i\upsilon_j-\frac{Gm}{r}n_in_j),
\end{equation}
here, $M$ is the mass of the black hole, $m$ is the mass of the test particle, $\mathbf{\upsilon}$ is the relative velocity, $\mathbf{n}$ is the direction of the separation vector of the EMRI system, $\eta = \frac{Mm}{{(M+m)}^2}$ is the symmetric mass ratio, and $D_L$ is the luminosity distance from the EMRI system to the detector.

To construct the gravitational wave waveform, we introduce a detector adaptive coordinate system $(X, Y, Z)$ outside the original coordinate system $(x, y, z)$\cite{Will:2016sgx}, and its coordinate directions are
\begin{equation}\label{31}
\begin{split}
&\mathbf{e}_X=\left(\cos{\zeta},-\sin{\zeta},0\right),\\& \mathbf{e}_Y=\left(\cos{\iota}\sin{\zeta},\cos{\iota}\cos{\zeta},-\sin{\iota}\right),\\& \mathbf{e}_Z=(\sin{\iota}\sin{\zeta},\sin{\iota}\cos{\zeta},\cos{\iota}).
\end{split}
\end{equation}
Here, using $\mathbf{e}_X$ and $\mathbf{e}_Y$ as vector bases, there exists
\begin{equation}\label{32}
\begin{split}
&h_+=\frac{1}{2}(e_X^ie_X^j-e_Y^ie_Y^j)h_{ij},\\&h_\times=\frac{1}{2}(e_X^ie_Y^j+e_Y^ie_X^j)h_{ij}.
\end{split}
\end{equation}
According to Eqs. \eqref{30}-\eqref{32}, the corresponding gravitational wave is obtained\cite{Will:2016sgx}.
\begin{equation}\label{33}
h_+=-\frac{2\eta}{c^4D_L}\frac{\left(GM\right)^2}{r}\left(1+\cos^2{\iota}\right)\cos{\left(2\phi+2\zeta\right)},
\end{equation}
\begin{equation}\label{34}
h_\times=-\frac{4\eta}{c^4D_L}\frac{\left(GM\right)^2}{r}\cos{\iota}\sin{\left(2\phi+2\zeta\right)}.
\end{equation}
Among them, $\iota$ is the inclination angle between the orbital plane of the test particle and the $X-Y$ plane in the detector coordinate system, $\zeta$ is the latitude angle and $\phi$ is the phase angle.

\begin{figure*}[]
\includegraphics[width=1\textwidth]{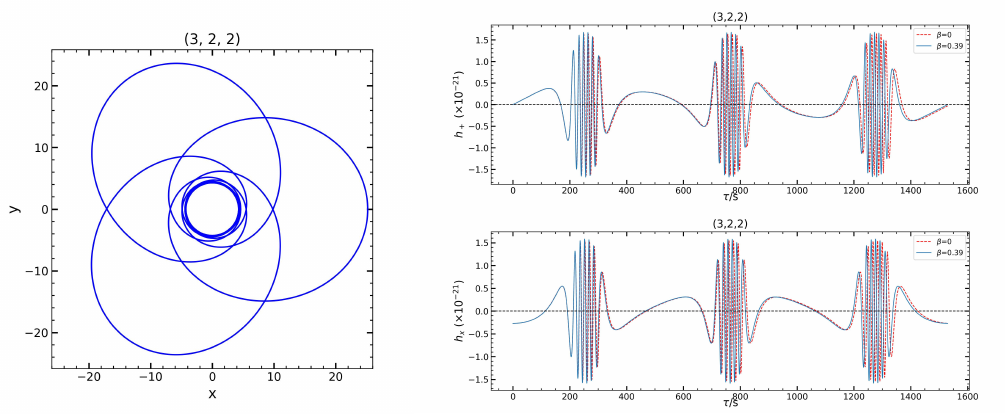}
\caption{The left side shows the trajectory of the periodic orbit $(z,w,v)=(3,2,2)$. The right side features a red dashed line representing the gravitational wave of a Schwarzschild black hole, with parameters set to $\beta=0$ and $q=2+\frac{2}{3}$. The light blue solid line represents the gravitational wave of a regular hairy black hole, with parameters set to $\beta=0.39$ and $q=2+\frac{2}{3}$.}
\label{j}
\end{figure*}

\begin{figure*}[]
\includegraphics[width=1\textwidth]{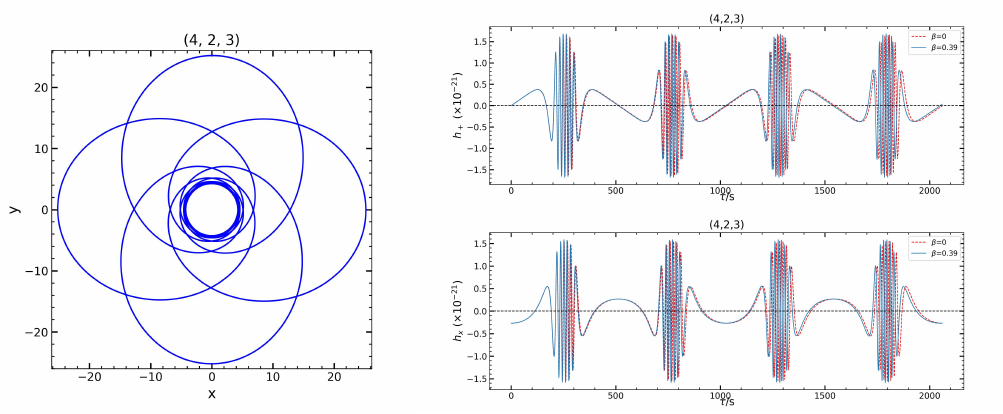}
\caption{The left side shows the trajectory of the periodic orbit $(z,w,v)=(4,2,3)$. The right side features a red dashed line representing the gravitational wave of a Schwarzschild black hole, with parameters set to $\beta=0$ and $q=2+\frac{3}{4}$. The light blue solid line represents the gravitational wave of a regular hairy black hole, with parameters set to $\beta=0.39$ and $q=2+\frac{3}{4}$.}
\label{k}
\end{figure*}

To plot the corresponding gravitational waveforms, we consider the gravitational waves radiated from a complete periodic orbit and set the parameters of the EMRI system as $\iota=\frac{\pi}{4}$, $\zeta=\frac{\pi}{4}$, $M={10}^7M_\odot$, and $m=10M_\odot$, where $M_\odot$ is the solar mass. The luminosity distance is $D_L=200$ $Mpc$. Next, in Figs.\ref{j} and \ref{k}, we consider two classical periodic orbits: (3,2,2) and (4,2,3). As shown in Figs.\ref{j} and \ref{k}, the gravitational wave exhibits a distinct zoom-whirl behavior over a complete orbit. Combined with the orbital period diagram on the left, it can be seen that during the zoom phase, where the orbit is highly elliptical, the gravitational wave waveform changes relatively smoothly; whereas during the whirl phase, where the orbit becomes nearly circular, the gravitational wave waveform undergoes significant variations. These variations in the gravitational wave waveform correspond to the orbital precession (the number of orbital leaves), indicating that the gravitational wave signal reflects the periodic orbital characteristics of the system. 

In addition, compared with the Schwarzschild black hole ($\beta = 0$), the existence of the hair parameter $\beta$ has made subtle corrections to the gravitational wave waveform. Specifically, it is manifested as an increase in the amplitude of the gravitational wave, a change in the period, and a change in the phase. Although these differences are not significant in a single periodic orbit, during the inspiral phase of an extreme-mass-ratio inspiral (EMRI) system, the companion star typically orbits the central black hole $10^3$ to $10^5$ times, continuously dissipating the system's orbital angular momentum $L$ and energy $E$ through gravitational wave radiation. This process causes the particle to gradually migrate to periodic orbits closer to the black hole. During this long-term evolution, the phase shift, amplitude variation, and period correction of the gravitational waves gradually amplify through cumulative effects, resulting in significant waveform deviations over extended evolution. This cumulative effect not only allows precise gravitational wave detection to reveal the influence of the hair parameter on the dynamics of matter near black holes but also provides a unique observational window for exploring the multi-hair structure of black holes. In the future, we will focus on constructing more precise gravitational wave waveforms while fully considering the effects of other matter fields on gravitational waves. By accurately modeling and thoroughly analyzing these effects, it is expected that differences between the regular hairy black hole and the Schwarzschild black hole can be detected through the observation of gravitational wave signals emitted during long-term orbital evolution. This will provide stronger theoretical support and observational evidence for testing the no-hair theorem and constraining the subtle features of the gravitational field of black holes.

It is worth noting that the gravitational wave waveform obtained for a single periodic orbit is not an exact waveform. This is because, in our calculations, we adopted the adiabatic approximation, performed the analysis within the framework of pure general relativity, and used the quadrupole formula (equation \eqref{30}), neglecting the contributions from second-order and higher-order multipole moments. Nevertheless, our calculation results not only reflect the fundamental characteristics of gravitational wave radiation from periodic orbits but also reveal the subtle influence of the hair parameter on gravitational wave signals. Although this influence is relatively minor within a single period, it accumulates over long-term orbital evolution, potentially leading to significant effects on observations. Current research indicates that analyzing the evolution of periodic orbits in EMRI systems holds significant physical importance\cite{Mino:1996nk}. In addition, the EMRI system is considered as one of the key signal sources in gravitational wave observations\cite{Amaro-Seoane:2007osp,Babak:2017tow}. The study of this system not only helps to deeply understand the nature of black holes\cite{Amaro-Seoane:2007osp} but also can help explore whether there is dark matter around black holes\cite{Hannuksela:2019vip} and impose more precise constraints on cosmological parameters\cite{Schutz:1986gp}. In summary, EMRI systems hold tremendous research value in gravitational wave astronomy and serve as an important window for understanding black hole physics and cosmology.

\section{Summary and discussion}\label{5.0}
This paper systematically analyzes the timelike geodesic properties of the regular hairy black hole and investigates the impact of the hair parameter on the event horizon, bound orbits, and gravitational wave radiation. Our analysis shows that the hair parameter has a significant impact on the spacetime structure near the event horizon. The presence of the hair parameter causes the event horizon radius of a regular hairy black hole to be smaller than that of a Schwarzschild black hole. In the further analysis of the relationship between the marginally bound orbit (MBO) and the innermost stable circular orbit (ISCO) parameters and the hair parameter $\beta$, the results show that when the hair parameter is small, the changes in the parameters of the MBO and ISCO are not significant, indicating that it is difficult to distinguish between the regular hairy black hole and the Schwarzschild black hole in this case. However, as the hair parameter $\beta$ increases, $r_{MBO}$, $L_{MBO}$, $r_{ISCO}$, $L_{ISCO}$, and $E_{ISCO}$ all exhibit a significant decreasing trend. Subsequently, through the analysis of the orbital angular momentum $L$ and energy $E$, we found an allowed parameter space. As the hair parameter $\beta$ increases, the allowed region gradually expands and shifts toward lower angular momentum and higher energy.

Further analysis reveals that, for the same periodic orbit, compared to the Schwarzschild black hole, the periapsis radius of the test particle in a regular hairy black hole is slightly closer to the black hole, while the apoapsis radius is slightly farther from the black hole. Although these differences are numerically small (see Tables \ref{table1} and \ref{table2}), they reflect the subtle modifications to the spacetime geometry of the black hole introduced by the hair parameter $\beta$. Specifically, this modification allows the test particle to enter stable periodic orbits with lower energy or angular momentum, thereby revealing the unique dynamical characteristics of the regular hairy black hole. 

In addition, by simulating the precession observations of the S2 star orbiting Sgr A*, the hair parameter was constrained. The results indicate that within the parameter range where a black hole exists ($0 \leq \beta \leq 0.3906$), the correction effect of the hair parameter satisfies the current observational constraints, but no stricter constraints could be established. Furthermore, we explored the periodic orbit behavior of the regular hairy black hole. Research shows that the rational number parameter $q$ of the periodic orbit increases with the increase of orbital energy and decreases with the increase of orbital angular momentum. Further analysis shows that as the hair parameter $\beta$ increases, the extreme values of orbital angular momentum and energy both show a downward trend (see Fig.\ref{g}). In addition, through numerical analysis of different periodic orbits $(z, w, v)$, we come to the conclusion that for the same periodic orbit, a regular hairy black hole has lower angular momentum $L$ and energy $E$ than a Schwarzschild black hole (see tables \ref{table1} and \ref{table2}). Further analysis from the periodic trajectory shows that the test particle around a regular hairy black hole can remain stable on a periodic orbit closer to the black hole. It is speculated that this phenomenon may be related to the fact that the hair parameter $\beta$ changes the spacetime structure of the black hole. Although the influence of the hair parameter $\beta$ is relatively weak, these subtle corrections play a significant role in enhancing orbital stability and may manifest as observable features in gravitational wave signals through cumulative effects during long-term orbital evolution. Such modifications in the spacetime of the regular hairy black hole not only provide the possibility of exploring the existence of black hole hair parameters but also offer new theoretical support for high-precision gravitational wave observations.

Finally, treating periodic orbits as transitional orbits in an extreme-mass-ratio inspiral (EMRI) system, we studied the gravitational wave radiation characteristics of a single period. The results show that, over a complete orbital period, the gravitational wave signal clearly exhibits the zoom-whirl behavior of periodic orbits. Additionally, the hair parameter $\beta$ induces subtle effects on the phase, amplitude, and period of the gravitational wave waveform. Although these effects are relatively weak within a single period, they may accumulate into significant effects over long-term evolution, leading to notable deviations in gravitational wave waveforms during extended evolution. This finding suggests that future high-sensitivity gravitational wave detectors (such as LISA, Taiji, and TianQin) have the potential to detect the influence of the hair parameter, thereby deepening our understanding of the spacetime properties and dynamical behavior around black holes. This not only helps to distinguish between a Schwarzschild black hole and a regular hairy black hole, thereby testing the validity of the no-hair theorem, but also provides new theoretical support and potential advantages for high-precision measurements in gravitational wave detection.

Overall, this study reveals the impact characteristics of the hair parameter in the regular hairy black hole on different regions and dynamical phenomena. It indicates that its dynamical effects are relatively weak near bound orbits but exhibit significant effects near the event horizon, with potential observable features emerging during long-term gravitational wave evolution. Therefore, further in-depth exploration of the properties and effects of the hair parameter requires high-sensitivity observational tools, such as space-based gravitational wave detectors (e.g., LISA, Taiji, and TianQin) and black hole shadow observations. This study provides a theoretical basis for understanding the dynamical properties and physical significance of the regular hairy black hole and may offer new approaches for future observations and analyses under strong gravitational field conditions. It is worth noting that this paper only considers gravitational wave radiation within a single period, which is sufficient for evaluating the observational value of the hair parameter. However, when analyzing gravitational wave signals from long-term orbital evolution in the future, it will be necessary to further account for the back-reaction of gravitational radiation on the orbit. In addition, in the real cosmic environment, black holes are usually rotating, and their surroundings are filled with matter. Therefore, environmental effects on the orbits also require special attention. At the same time, since the gravitational wave signals from the extreme-mass-ratio inspiral (EMRI) system are relatively weak, detecting such signals requires long-term accumulation to achieve a sufficient signal-to-noise ratio. Therefore, the design of waveform templates must ensure both high precision and high efficiency. These are key challenges that need to be addressed in future research.

\section*{Acknowledgements}
We acknowledge the anonymous referee for a constructive report that has significantly improved this paper. This work was supported by Guizhou Provincial Basic Research Program (Natural Science) (Grant No. QianKeHeJiChu-[2024]Young166), the Special Natural Science Fund of Guizhou University (Grant No.X2022133), the National Natural Science Foundation of China (Grant No. 12365008) and the Guizhou Provincial Basic Research Program (Natural Science) (Grant No. QianKeHeJiChu-ZK[2024]YiBan027).

\section*{Appendix}
In this appendix, we will briefly outline the derivation process of the metric \eqref{2}, explaining how the metric is derived from the Einstein-Hilbert action through the gravitational decoupling method and satisfies the Einstein field equations, while also providing the corresponding Lagrangian density.

Jorge Ovalle and other scholars adopted the gravitational decoupling method, starting from the Einstein-Hilbert action, which is expressed as follows\cite{Ovalle:2023ref}
\begin{equation}\label{35}
S=\int{\left[\frac{R}{2\kappa}+\mathcal{L}_M+\mathcal{L}_\Theta\right]\sqrt{-g}}d^4x,
\end{equation}
here, $ R $ is the Ricci scalar, $ \mathcal{L}_M $ is the Lagrangian density containing the standard matter, and $ \mathcal{L}_\Theta $ is the Lagrangian density introduced for additional sources or gravitational sectors beyond general relativity.

For these two sources, their energy-momentum tensors can be written as
\begin{equation}\label{36}
T_{\mu\nu}=-\frac{2}{\sqrt{-g}}\frac{\delta\left(\sqrt{-g}\mathcal{L}_{M}\right)}{\delta g^{\mu\nu}}=-2\frac{\delta\mathcal{L}_{M}}{\delta g^{\mu\nu}}+g_{\mu\nu\mathcal{L}_{M}},
\end{equation}
\begin{equation}\label{37}
\theta_{\mu\nu}=-\frac{2}{\sqrt{-g}}\frac{\delta\left(\sqrt{-g}\mathcal{L}_\Theta\right)}{\delta g^{\mu\nu}}=-2\frac{\delta\mathcal{L}_\Theta}{\delta g^{\mu\nu}}+g_{\mu\nu\mathcal{L}_\Theta}.
\end{equation}
At this point, the Einstein field equations can be written as
\begin{equation}\label{38}
G_{\mu\nu}\equiv R_{\mu\nu}-\frac{1}{2}Rg_{\mu\nu}=\kappa\left(T_{\mu\nu}+\theta_{\mu\nu}\right),
\end{equation}
here $\kappa=8\pi G$.

For a static spherically symmetric solution, it can typically be expressed as
\begin{equation}\label{39}
ds^2=-e^{\nu\left(r\right)}dt^2+e^{\lambda\left(r\right)}dr^2+r^2d\Omega^2.
\end{equation}
Substituting the above expression into the field equation \eqref{38} yields the corresponding energy-momentum tensor
\begin{equation}\label{40}
T_\mu^\nu=\mathrm{diag}\left[-\varepsilon,p_r,p_\theta,p_\theta\right],
\end{equation}
\begin{equation}\label{41}
\theta_\mu^\nu=\mathrm{diag}\left[-\mathcal{E},\mathcal{P}_{r},\mathcal{P}_\theta,\mathcal{P}_\theta\right].
\end{equation}
The detailed derivation process can be found in the original reference\cite{Ovalle:2023ref}.

In order to obtain the regular hairy black hole solution, we use the traditional black hole solution as a seed source, and on this basis, introduce additional matter sources to avoid the emergence of singularities. The expression for the seed source is
\begin{equation}\label{42}
ds^2=-e^{\xi\left(r\right)}dt^2+e^{\mu\left(r\right)}dr^2+r^2d\Omega^2.
\end{equation}
After adding the additional source $ \theta_{\mu\nu} $, the corresponding function can be written as
\begin{equation}\label{43}
\xi\rightarrow\nu=\ \xi+\ \alpha g,
\end{equation}
\begin{equation}\label{44}
e^{-\mu}\rightarrow e^{-\lambda}=e^{-\mu}+\alpha f.
\end{equation}
Here, $ \alpha $ is the deformation parameter, and $ g$ and $ f$ are correction functions related to the matter source.

Combining expressions \eqref{43} and \eqref{44}, the Einstein equation \eqref{38} can be split into two sets. The first set is given by $ T_{\mu\nu} $, and at this point, we have
\begin{equation}\label{45}
\begin{split}
&\kappa\varepsilon=\frac{1}{r^2}-e^{-\mu}\left(\frac{1}{r^2}-\frac{\mu^\prime}{r}\right),\\&
\kappa p_r=-\frac{1}{r^2}+e^{-\mu}\left(\frac{1}{r^2}+\frac{\xi^\prime}{r}\right),\\&
\kappa p_\theta=\frac{e^{-\mu}}{4}\left(2\xi^{\prime\prime}+\xi^{\prime2}-\mu^\prime\xi^\prime+2\frac{\xi^\prime-\mu^\prime}{r}\right).
\end{split}
\end{equation}
The second set is given by the source $ \theta_{\mu\nu} $, and at this point, we have
\begin{equation}\label{46}
\begin{split}
&\kappa\mathcal{E}=-\frac{\alpha f}{r^2}-\frac{\alpha f^\prime}{r},\\&
\kappa\mathcal{P}_{r}-\alpha Z_1=\alpha f\left(\frac{1}{r^2}+\frac{\nu^\prime}{r}\right),\\&
\kappa\mathcal{P}_\theta-\alpha Z_2=\frac{\alpha f}{4}\left(2\nu^{\prime\prime}+\nu^{\prime2}+2\frac{\nu^\prime}{r}\right)-\frac{\alpha f^{\prime}}{4}\left(\nu^\prime+\frac{2}{r}\right).
\end{split}
\end{equation}
where
\begin{equation}\label{47}
Z_1=\frac{e^{-\mu}g^\prime}{r},
\end{equation}
\begin{equation}\label{48}
4Z_2=e^{-\mu}\left(2g^{\prime\prime}+\alpha g^{\prime2}+\frac{2g^\prime}{r}+2g^\prime\xi^\prime-\mu^\prime g^\prime\right).
\end{equation}
The energy exchange between these two sources is given by the conservation equation
\begin{equation}\label{49}
\nabla_\sigma T_\nu^\sigma=-\frac{\alpha g^\prime}{2}\left(\varepsilon+p_r\right)\delta_\nu^\sigma=-\nabla_\sigma\theta_\nu^\sigma.
\end{equation}

In order to construct a regular hairy black hole solution within the framework of general relativity, the authors of the original paper use the Schwarzschild black hole solution ($ T_{\mu\nu} $) as a seed source and prevent the formation of singularities due to gravitational collapse by introducing an additional matter source $ \theta_{\mu\nu} $, such as the introduction of a tensor vacuum source. On this basis, they further require that the solution has a well-defined event horizon structure and satisfies the weak energy condition.

By applying these constraint conditions, the regular hairy black hole solution is ultimately obtained\cite{Ovalle:2023ref}
\begin{equation}\label{50}
e^\nu=e^{-\lambda}=1-\frac{2M}{r}+\frac{e^{-\alpha r/M}}{rM}\left(\alpha^2r^2+2M\alpha r+2M^2\right).
\end{equation}
This metric describes the regular black hole solution after the deformation is introduced through gravitational decoupling and the addition of sources. The hair parameter $ \alpha $ controls the structure of the metric. As $ \alpha \to 0 $, the metric degenerates into Minkowski spacetime, and as $ \alpha \to \infty $, the metric recovers the standard Schwarzschild solution.

In addition, the metric \eqref{50} can also be derived through the variational principle from the assumed action. The total Lagrangian density $ \mathcal{L} $ (which includes the additional matter source) can be obtained using the P-dual formalism\cite{Ayon-Beato:1998hmi}. In this approach, the expression for the total action is
\begin{equation}\label{51}
S=\int{\left[\frac{R}{2\kappa}+\mathcal{L}\right]\sqrt{-g}}d^4x.
\end{equation}
The specific form of the Lagrangian density $ \mathcal{L} $ is\cite{Ovalle:2023ref}
\begin{equation}\label{52}
\mathcal{L}\left(P\right)=\frac{\alpha^3}{2\kappa{M}^3}\left[\xi\left(P\right)-2{M}\right]e^{-\xi\left(P\right)/{M}},
\end{equation}
here
\begin{equation}\label{53}
\xi\left(P\right)=\alpha\left(\frac{-2\alpha^2}{\kappa^2P}\right)^\frac{1}{4}.
\end{equation}
The detailed derivation of this part can be found in the reference \cite{Ovalle:2023ref}.

\bibliography{ref}
\bibliographystyle{apsrev4-1}

\end{document}